\documentclass[preprint,12pt,authoryear]{elsarticle}
\usepackage{graphics,bm,graphicx}
\usepackage{float}
\usepackage[top=1.25in, bottom=1.25in, left=1.25in, right=1.25in]{geometry}
\usepackage{tikz}
\usepackage{pdfpages}
\usepackage{graphicx}
\usepackage{grffile}
\usepackage{bm}
\usepackage{multirow}
\usepackage[export]{adjustbox}

\usepackage[subrefformat=parens,labelformat=parens]{subfig}
\usepackage{mathtools}
\usepackage{amsmath,mathptmx}
\usepackage[displaymath,mathlines]{lineno}
\usepackage{amssymb}
\usepackage{multicol}
\usepackage{booktabs}

\usepackage{rotating}
\usepackage{natbib}
\setcitestyle{authoryear,open={(},close={)}} 

\begin{document}
\begin{frontmatter}

\title{Inner core heterogeneity induced by a large variation in
lower mantle heat flux}
\author{Aditya Varma, Binod Sreenivasan }
\address{Centre for Earth Sciences, Indian Institute of
Science, Bangalore 560012, India}
\date{}

\begin{abstract}
Seismic mapping of the top of the inner core
indicates two distinct areas of high P-wave velocity,
the stronger one located beneath Asia, and
	the other located beneath the Atlantic. This 
	two-fold pattern supports the idea that a lower-mantle
	heterogeneity can be transmitted to the inner core
	through outer core convection. In this study,
a two-component convective dynamo model, where thermal
convection is near critical and compositional convection is
strongly supercritical, produces a substantial inner core heterogeneity 
in the rapidly rotating strongly driven regime of Earth's core.
While the temperature profile that models secular cooling
ensures that the mantle heterogeneity propagates as far as
the inner core boundary (ICB),
 the distribution of heat flux at the ICB
is determined by the strength of compositional buoyancy. A large
heat flux variation $q^*$ of $O(10)$ at the core--mantle
boundary (CMB), where $q^*$ is the ratio of the maximum heat flux
difference to the mean heat flux at the CMB, produces a
core flow regime of long-lived convection in the east
	 and time-varying convection in the west. Here,
	 the
	 P-wave velocity estimated from the ICB
	 heat flux in the dynamo is higher in the
	 east than in the west,
	 with the hemispherical difference of the same order
	 as the observed lower bound, 0.5 \%. 
	 Additional observational constraints are satisfied
	 in this regime -- the variability
	 of high-latitude magnetic flux in the east is 
	 lower than that in the west; 
	 and the stratified F-layer at the base of the outer core,
	 which is fed by the mass flux from
	 regional melting of the inner core 
	 and magnetically damped, attains a steady-state
	 height of $\sim$ 200 km.

\end{abstract}

\end{frontmatter}
	
	\section{Introduction}
	\label{intro}
	
	Seismic tomography of the lower mantle indicates
	regions of high shear wave velocity beneath
	Asia and the Atlantic and low velocity beneath the Pacific
	and Africa \citep[e.g][]{masters2000}. 
	While the fast regions cause preferential cooling
	of the outer core, the slow regions inhibit core
	convection. The high-latitude lobes of magnetic
	flux beneath Asia and America and the weak secular
	variation in the Pacific point to the effect of
	the approximately two-fold pattern of lower-mantle
	heterogeneity on the geodynamo \citep{willis2007}.
	 Since the heat flux variation at the core--mantle
	 boundary (CMB) is not likely to have changed much in
	 the past $\sim$ 200 Myr \citep{torsvik2006}, the evolution
	 of the geodynamo subject to a quasi-stationary
	  lateral variation in
	 lower-mantle heat flux is reproduced by numerical
	 dynamo simulations \citep[e.g.][]
	 {takahashi2008,olson2015,mound2023}. 
	 Although
	a simplified two-fold pattern of convection in the 
	core is often used to explain the non-axisymmetry
	of the magnetic field induced by the mantle, a large
	variation in CMB heat flux 
	\citep{olson2018,mound2019} gives rise
	to a substantial east--west
	dichotomy in the core flow \citep{sahoo2020a},
	consisting of coherent (long-lived) convection
	beneath Asia and time-varying convection beneath
	America. If this pattern of convection in turn causes
	a heterogeneity at the inner
	core boundary (ICB), then a measure of this
	heterogeneity would place
	a constraint on the magnitude of the heat flux
	variation at the CMB.
	
	Seismic mapping of the top of the inner core indicates
	an east--west lateral
	heterogeneity, with the East having
	a higher isotropic velocity than the West.
	The differences in the isotropic P-wave velocity $v_p$ 
	at the top of the inner	core lie between
	 $0.5\%$ and $1.5\%$ 
	\citep{tanaka1997,niu2001,
	sun2008,waszek2011}. 
	Isotropy, in this context, means that 
	the wave speeds are the same in all directions.
	Later studies suggest a more complex velocity structure,
	with the east--west
	variation showing a dominant two-fold (wavenumber $m=2$) pattern 
 {\citep{cormier2013,tkalvcic2024} -- there are two distinct areas of high
	velocity in the upper inner core, 
	the stronger one located beneath Asia, and
	the other located beneath the Atlantic. This pattern
	lends credence to the idea that a large lower-mantle
	heterogeneity can be transmitted to the inner core
	via outer core convection.
	
	Numerical dynamo simulations 
	that study the influence of the
	laterally varying
	lower-mantle heat flux on
	the geodynamo 
	suggest that a cyclonic vortex beneath
	Asia causes a hemispherical variation
	in the freezing rate of the inner core,
	with the eastern hemisphere freezing
	faster than the western hemisphere
	\citep{aubert2008,gubbins2011}.
	In addition, models that simulate
	secular cooling of the core through  
	volumetric heat sources produce areas of
	reversed (inward) heat flux at the ICB, 
	indicating possible
	 melting of the inner core \citep{gubbins2011}.
	Laboratory models of the core 
	where convection onsets
	in close proximity to the imposed lateral variation
	\citep{sahoo2020b} 
	 also produce heat flux
	reversals at the ICB. In contrast, basally heated
	numerical models operating in the strongly forced
	regime of present-day Earth \citep{sreenivasan2008}
	swamp any influence of the CMB heat flux anomalies,
	which might lead one to infer that
	the CMB anomalies do not reach the ICB \citep{davies2019}.
	It may be noted that similar models, where
	the maximum heat flux variation at the CMB is $\sim$ 5
	times the mean heat flux \citep{sahoo2020a,mound2023}, 
	reproduce the high-latitude
	magnetic flux in the present-day field well. To
	satisfy the observational constraints placed by
	the field as well as the heterogeneity
	of the ICB, the present study considers a two-component
	convective dynamo model. Here, the strong compositional buoyancy
	ensures the east--west variation of the observed field
	while the relatively weak thermal buoyancy that mimics
	secular cooling ensures that the mantle-induced anomalies
	not only reach the ICB but also provide a measurable
	estimate of the seismic velocity difference at the top
	of the inner core.

    Seismic observations suggest the presence of a
    low-$v_p$ layer at the base of the outer core. This 
    stably stratified \lq F-layer' 
    \citep{souriau1991,gubbins2008} may form
    from the lateral spread of melt originating from the
    regions of inward heat flux at the ICB. By varying the lower-mantle
    heat flux variation $q^*$ in the range $\sim$ 1--10 in
    dynamo models, where
    $q^*$ is the ratio of the maximum heat flux variation
    to the mean heat flux at the CMB, the lowest $q^*$ at
    which heat flux reverses at the ICB may be found. Similar to
    the pattern of inner core freezing, we anticipate a 
    two-fold pattern for melting, with an area beneath
    the Pacific being
    more dominant than that beneath Africa. 
    The height of the F-layer at the equator is a function
    of the
    mean hemispherical heat flux and the axial
    ($z$) field intensity
    in the dynamo. The present study aims to show that
the F-layer heights predicted by observations can be realized
by inner core heat flux variations induced by lower-mantle
variations of $q^*$ $O(10)$.

	The paper is organized as follows. In Section 2,
 the dynamo model and the core flows
that would produce the heat flux heterogeneity at the ICB
are discussed. 
	Based on the east--west diffrences in the heat flux
measured in the dynamo simulations, estimates are obtained for the hemispherical
differences in inner core freezing rate and P-wave velocity, the comparison
of which with seismic observations would place a lower bound on $q^*$.
In addition, the simulations provide an estimate of the height of
stratified F-layer thought to be fed by regional melting of the inner core,
which also affords comparisons with 
predictions based on observations.
Further, the longitudinal variation of 
the high-latitude flux lobes in the
high-$q^*$ simulations are compared 
with that of the present-day field.
Section 3
gives the implications of this study for the Earth's core.

	\section{Inner core heterogeneity in a two-component
	convective dynamo model}\label{sec:ddmodel}

\subsection{The dynamo model}	
    We consider dynamo action in  
	a rotating, electrically conducting fluid confined
	between two concentric, co-rotating spherical surfaces 
	that correspond to the ICB and the
	CMB. The ratio of inner to outer
	radius, $r_i/r_o$, is chosen to be 0.35.
	To obtain the non-dimensional governing equations,
	lengths are scaled by
	the thickness of the spherical shell $L$, time is scaled by the
	magnetic diffusion time, $L^2/\eta$, where $\eta$ is the magnetic
	diffusivity, 
	the velocity field $u$ is scaled by $\eta/ L$, and
	the magnetic field $B$ is scaled by $(2\varOmega\mu \eta \rho)^{1/2}$ 
	where $\varOmega$ is the rotation rate, $\rho$ is the fluid
	density, and $\mu$ is the magnetic permeability. 
	The temperature $T$ is scaled by $\beta_T L$, 
where $\beta_T$ is the mean heat flux at
the CMB, and the composition $C$ is scaled by
$\beta_C L$, where $\beta_C$ is the mean composition flux at the ICB.
	
The governing equations for $\bm{u}$, $\bm{B}$, $T$ and $C$ are given by

\begin{linenomath}

	\begin{align}
E Pm^{-1}  \Bigl(\frac{\partial {\bm u}}{\partial t} + 
(\nabla \times {\bm u}) \times {\bm u}
\Bigr)+  {\hat{\bm{z}}} \times {\bm u} = - \nabla p^\star +
Pm Pr^{-1} E Ra^T T \bm{r} \,  \nonumber\\ +  
Pm Sc^{-1} E Ra^C C \bm{r} + (\nabla \times {\bm B})
\times {\bm B} + E\nabla^2 {\bm u}, \label{momentum} \\
\frac{\partial {\bm B}}{\partial t} = \nabla \times ({\bm u} \times {\bm B}) 
+ \nabla^2 {\bm B},  \label{induction}\\
\frac{\partial T}{\partial t} +({\bm u} \cdot \nabla) T =  Pm Pr^{-1} \,
\nabla^2 T +S_o,  \label{heat1}\\
\frac{\partial C}{\partial t} + (\bm{u} \cdot \mathbf \nabla) C= Pm Sc^{-1}\nabla^{2}C + S_{i}, \label{comp1} \\
\nabla \cdot {\bm u}  =  \nabla \cdot {\bm B} = 0,  \label{div}
\end{align}

\end{linenomath}

where the modified pressure $p^*$ is given by 
$p + \frac{1}{2} E Pm^{-1} |  \bm{u}^2|$.
The dimensionless parameters in the equations \eqref{momentum}--\eqref{div} 
are the Ekman number $E=\nu/2\varOmega L^2$,
the Prandtl number, $Pr=\nu/\kappa_T$, Schmidt number, $Sc= \nu/\kappa_C$,
the magnetic Prandtl number, $Pm=\nu/\eta$,
the thermal Rayleigh number 
$Ra^T= g \alpha \beta_T L^4/\nu \kappa_T$,
the compositional Rayleigh number $Ra^C= g \alpha \beta_C L^4/\nu \kappa_C$.
Here, $g$ is the gravitational 
acceleration, 
$\nu$ is the kinematic viscosity, 
$\kappa_T$ is the thermal diffusivity,  
$\kappa_C$ is the compositional diffusivity, 
and $\alpha$ 
is the coefficient of thermal expansion.
	
The velocity and magnetic fields satisfy the no-slip 
	and electrically insulating conditions, respectively.
The basic state temperature 
profile represents an internal heating 
with a source $S_o$, which models secular cooling \citep{willis2007}.
An isothermal condition is set at the ICB and a fixed
 heat flux condition at the CMB. The pattern of 
 heat flux at the CMB 
is derived from the seismic shear wave velocity
	variation in the lower mantle \citep{masters2000,
	sahoo2020a}.
The compositional profile uses a uniform volumetric sink
$S_i$ \citep[e.g.][]{manglik2010} with
a constant flux at the ICB and zero flux at the CMB. 
The basic state profiles of the temperature and the compositional gradient
are given by

\begin{linenomath}		
\begin{subequations}\label{tcprofiles}
\begin{gather}
T_0= \dfrac{Pr}{Pm} \, \dfrac{S_{o}}{6}(r_{i}^2-r^2), \quad 
\dfrac{\partial C_0}{\partial r}=  \dfrac{Sc}{Pm} \,\dfrac{S_{i}}{3} 
				\left( \dfrac{r_{o}^3}{ r^2}-r \right).
\tag{\theequation a,b}
\end{gather}
\end{subequations}
\end{linenomath}

	The fixed parameters used in the simulations are
	$Pr=0.1$, $Sc=1$ and $Pm=1$. The scaled magnetic
	field $\overline{B}$ is an output derived from our dynamo
	simulations as a root mean square (rms) value, where the mean is
	a volume average. 
	The input and output parameters of the simulations are given 
	in table \ref{tab:sim_param}. The calculations are performed by
	a code that uses spherical harmonic expansions in the
	angular coordinates $(\theta, \phi)$ and finite differences
	in radius $r$ \citep[e.g.][]{willis2007}.

	\begin{table}[!htbp]	
		\centering
		
		\begin{tabular}{|c c c c c c c c c c c|}
			\hline
			S. No. & $E$ & $q^*$ & $Ra^T$ & $Ra^{C} $& $Ra^{C}/Ra^C_{cr}$ & $N_{r}$ & $l$   & $\overline{B}$ & $Rm$ & $Ro_l$  \\
			& & &$(\times 10^6)$ & $(\times 10^9)$ & & & & & &  \\
			\hline
			a   & $2 \times 10^{-6}$  & 1     & 7.2    & 2     & 129                                    & 240      & 260 & 1.74                  & 567 & 0.010 \\
			b    &  & 2     & 6.75     & 2     & 129                                    & 256      & 260 & 1.68                  & 543  &0.012\\
			c    & & 7     &5.75 & 1         & 65                                      & 192      & 192 &
			1.22                  & 251  & 0.007\\
			d    & & 7     &5.75 & 1.5         & 97                                      & 192      & 230 &
			1.47                  & 405  & 0.010\\
			e     & & 7     & 5.75     & 2     & 129                                    & 240      & 260 & 1.65                  & 525 & 0.012  \\
			f     & & 7     & 5.75     & 2.75     & 177                                    & 240     & 288 & 1.95                 & 632 & 0.018  \\
			g     & & 10     & 4.3     & 2.75     & 177                                    & 256      & 270 & 2.05                  & 588 & 0.014  \\
			h     & & 10     & 4.3     & 3.50     & 225                                    & 256      & 288 & 2.22                  & 611 & 0.017 \\
			i     & & 10     & 4.3     & 4.35     & 275                                    & 240      & 288 & 2.41                  & 652 & 0.022\\
           j     & $1 \times 10^{-5}$ & 10     & 1.12    & 0.90     & 360                                 & 144      & 144 & 2.26                 & 372 & 0.023\\
            k     &                   & 15     & 0.9    & 1.05     & 420                                   & 144      & 160 & 2.47                 & 410 & 0.029\\
			\hline

		\end{tabular}
				\caption{Parameters of dynamo simulations in this study. Here, $N_{r}$ is 
			the number of radial points, $l$ is the spherical 
			harmonic degree, $\overline{B}$ is the volume-averaged
			magnetic field, $Rm$ is the magnetic 
			Reynolds number, $Ro_l$ is the local 
			Rossby number, $E$ is the Ekman number, 
			$Ra^T$ is the thermal Rayleigh number,
			$Ra^C$
			is the compositional Rayleigh number and $q^*$
			is the dimensionless measure of the 
			heat flux heterogeneity at the outer
			boundary. The common dimensionless parameters are
			$Pr=0.1$, $Sc=1$, $Pm=1$.
			}
		\label{tab:sim_param}
	\end{table}

	\subsection{Outer core flow and ICB heat flux}
\label{ichf}	
	
	In the simulations, the thermal Rayleigh number $Ra^T$ is 
	set to critical while the 
	compositional Rayleigh number $Ra^C$ is varied and is 
	kept highly supercritical, consistent with
	the respective estimates of $Ra/Ra_{cr}$ for
	nonmagnetic convection with turbulent
	diffusivities \citep[e.g.][]{gubbins2001}.
	The measure of lateral heterogeneity at the CMB, defined 
	by $q^{*}$ in Section \ref{intro}, 
	ranges from 1 to 15 in the simulations (table \ref{tab:sim_param}).

	We examine the core flow and ICB heat flux distribution 
	for varying $q^*$ and $Ra^C$. 
    The equatorial section
     and cylindrical surface plots of the radial 
    velocity $u_{r}$ are given in figure
    \ref{urplots}. The $(z,\phi)$ surface plots at
    cylindrical radius $s=1.2$, shown in figure 
    \ref{urplots} \subref{fig:urq7a}-\subref{fig:urq7c}, 
    indicate that the columnar structure 
    of the downwelling is preserved even at large $q^*$. Therefore,
    the dynamo operates in a rapidly rotating regime that
    generates a strong field of volume-averaged intensity
    $\overline{B} = O(1)$. For 
    $Ra^C \approx 65 \times Ra^C_{cr}$ and
    $q^{*}=7$, the convective downwelling beneath America 
    is dominant while the downwelling beneath 
	Asia is not well developed (figure \ref{urplots} 
	\subref{fig:urq7aeq} and 
	\subref{fig:urq7a}). This hemispherical dichotomy, found
	only at large $q^*$,
	has its origins in the onset of convection beneath the seismically
	faster region in the West \citep{sahoo2020a}. 
	As $Ra^C$ is increased to $\approx 100 \times Ra^C_{cr}$,
	coherent (long-lived) downwellings exist
	beneath both Asia and America  (figures 
	\ref{urplots}(b) and (e)), reminiscent of the convection
	pattern obtained with a $Y^2_{2}$ heat flux variation
	at the CMB \citep{willis2007}.
	 For a stronger forcing of $Ra^C \approx 130 \times Ra^C_{cr}$, 
	the downwelling beneath America disintegrates and
	gives way to a time-varying cluster of rolls
	 while the downwelling beneath Asia remains
	long-lived (figures \ref{urplots}(c) and (f)).
	The enhanced compositional convection beneath Asia
	might produce a thin low P-wave velocity layer 
	overlying a high-velocity, fast-freezing 
    region \citep[][p.~203]{cormier2021}.
	This core flow regime is likely Earth-like as it
produces high-latitude magnetic flux lobes which are more
unstable in the West than in the East in the historical period
of direct observation \citep{jackson2000,gubbins2007}.
	
	\begin{figure}[!htb]
		\centering 
		\subfloat[\label{fig:urq7aeq}]{
			\includegraphics[width=0.33\textwidth, keepaspectratio]{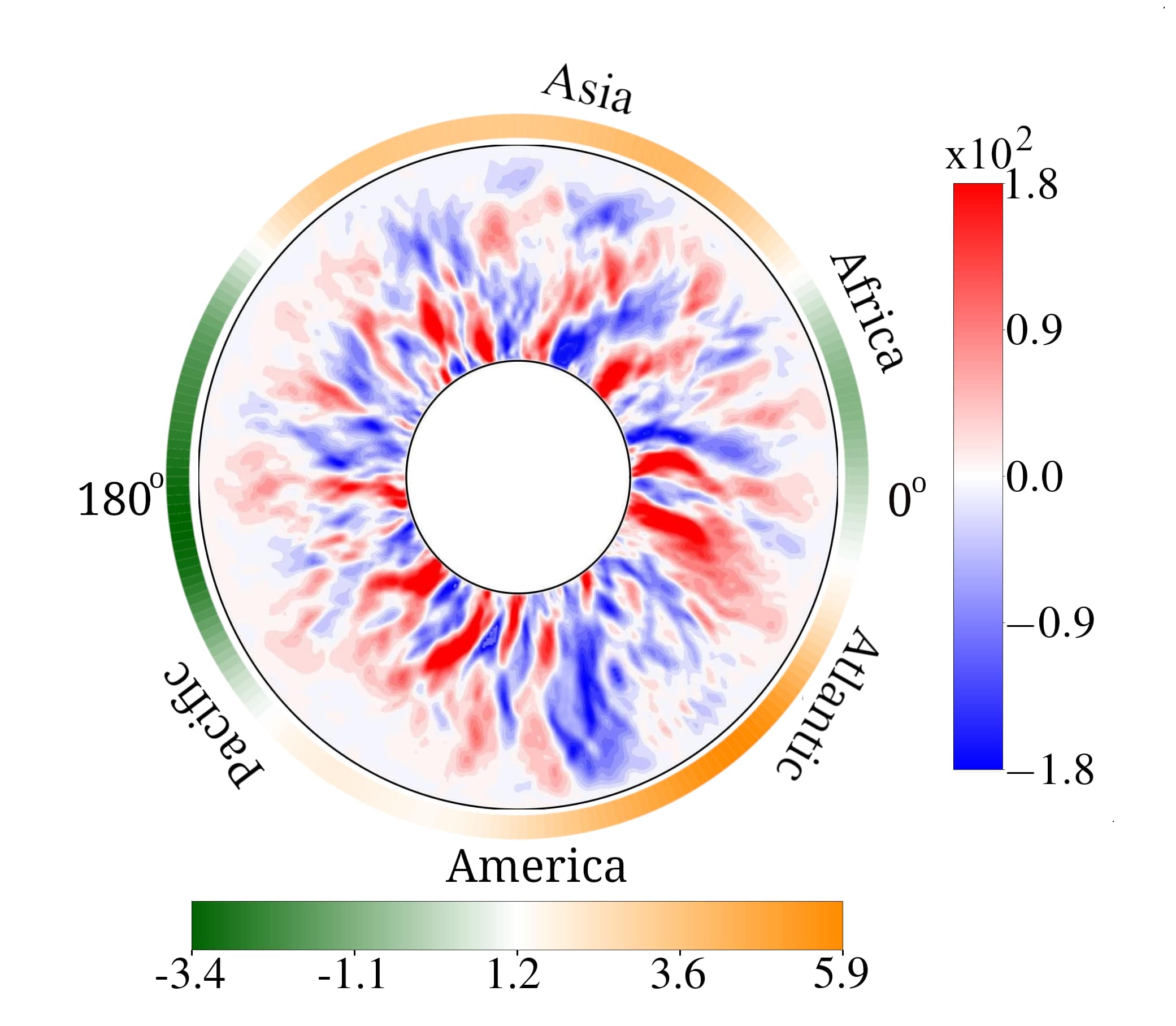}}	
			\subfloat[\label{fig:urq7beq}]{
			\includegraphics[width=0.33\textwidth, keepaspectratio]{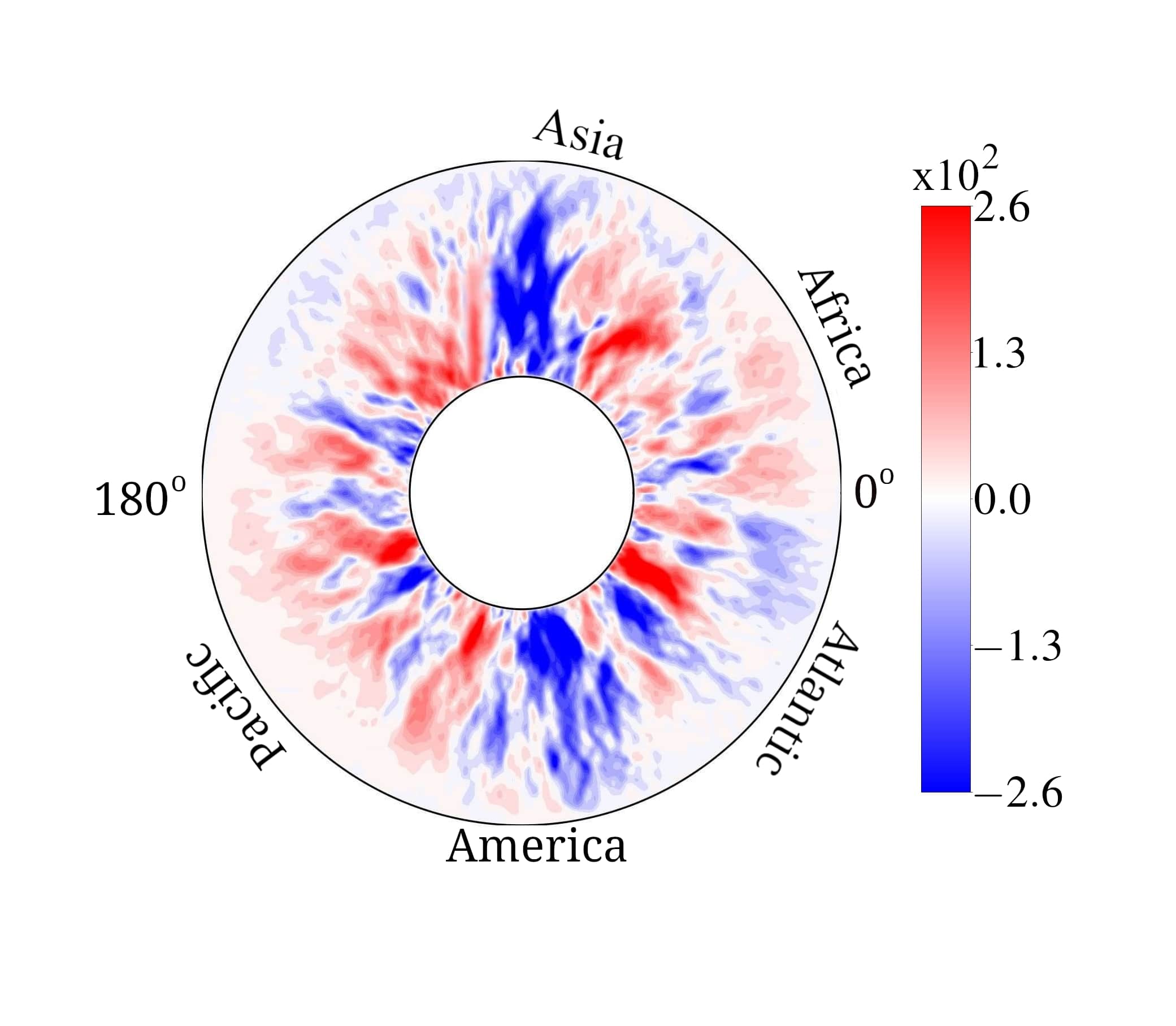}}	
			\subfloat[\label{fig:urq7ceq}]{
			\includegraphics[width=0.33\textwidth, keepaspectratio]{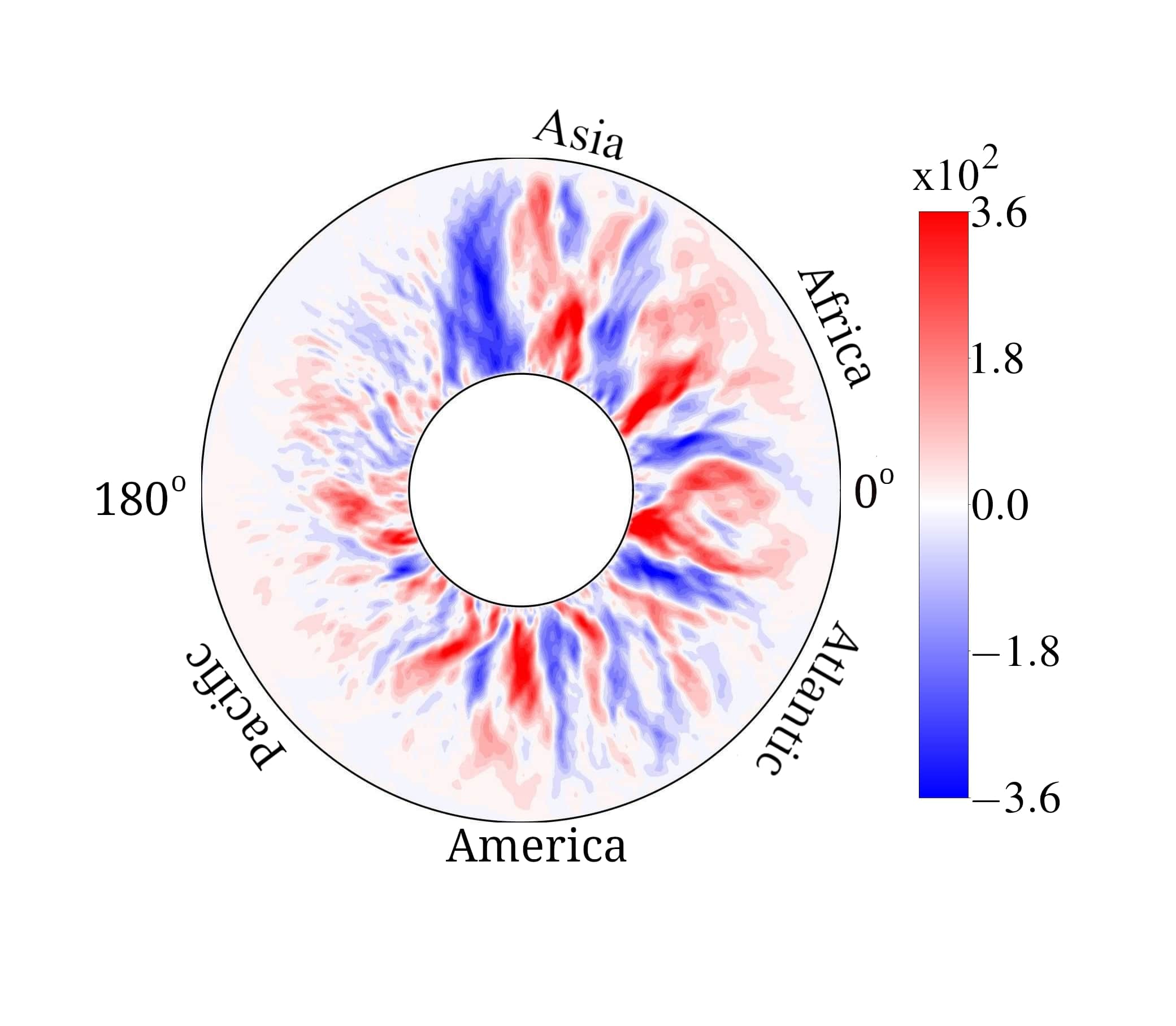}}	\\
		\subfloat[\label{fig:urq7a}]{
			\includegraphics[width=0.33\textwidth, keepaspectratio]{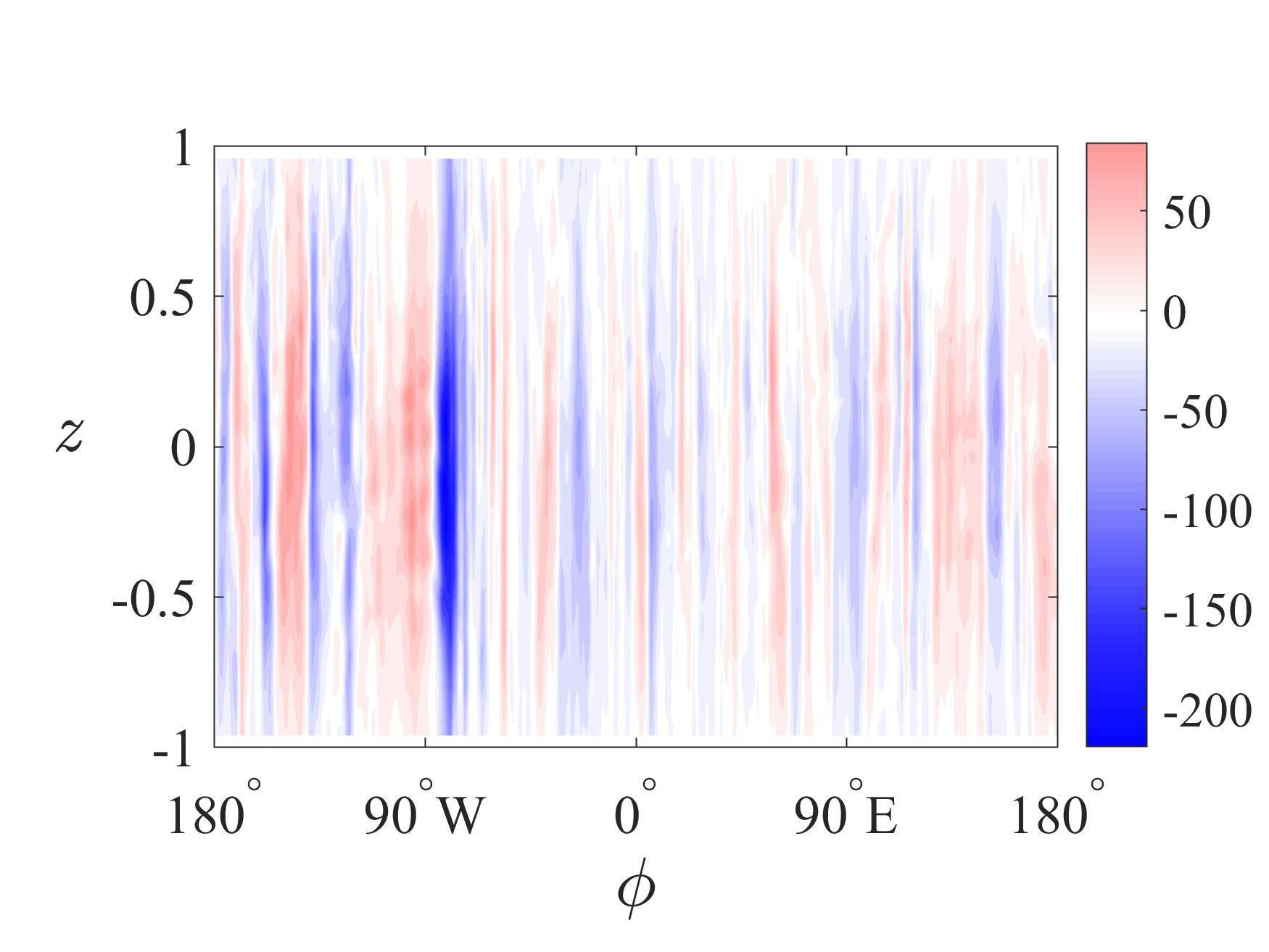}}	
		\subfloat[\label{fig:urq7b}]{
			\includegraphics[width=0.33\textwidth, keepaspectratio]{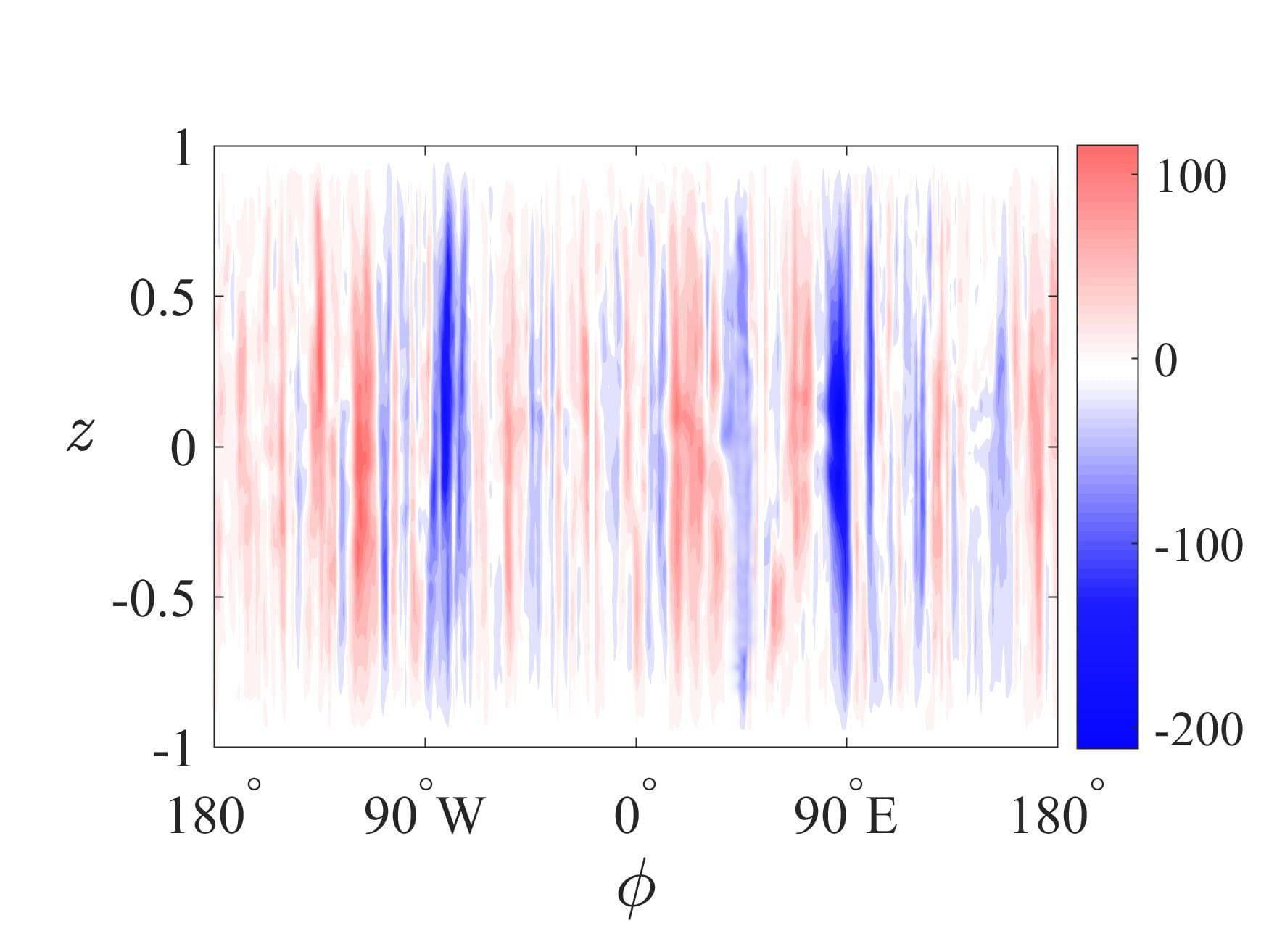}}
		\subfloat[\label{fig:urq7c}]{
			\includegraphics[width=0.33\textwidth, keepaspectratio]{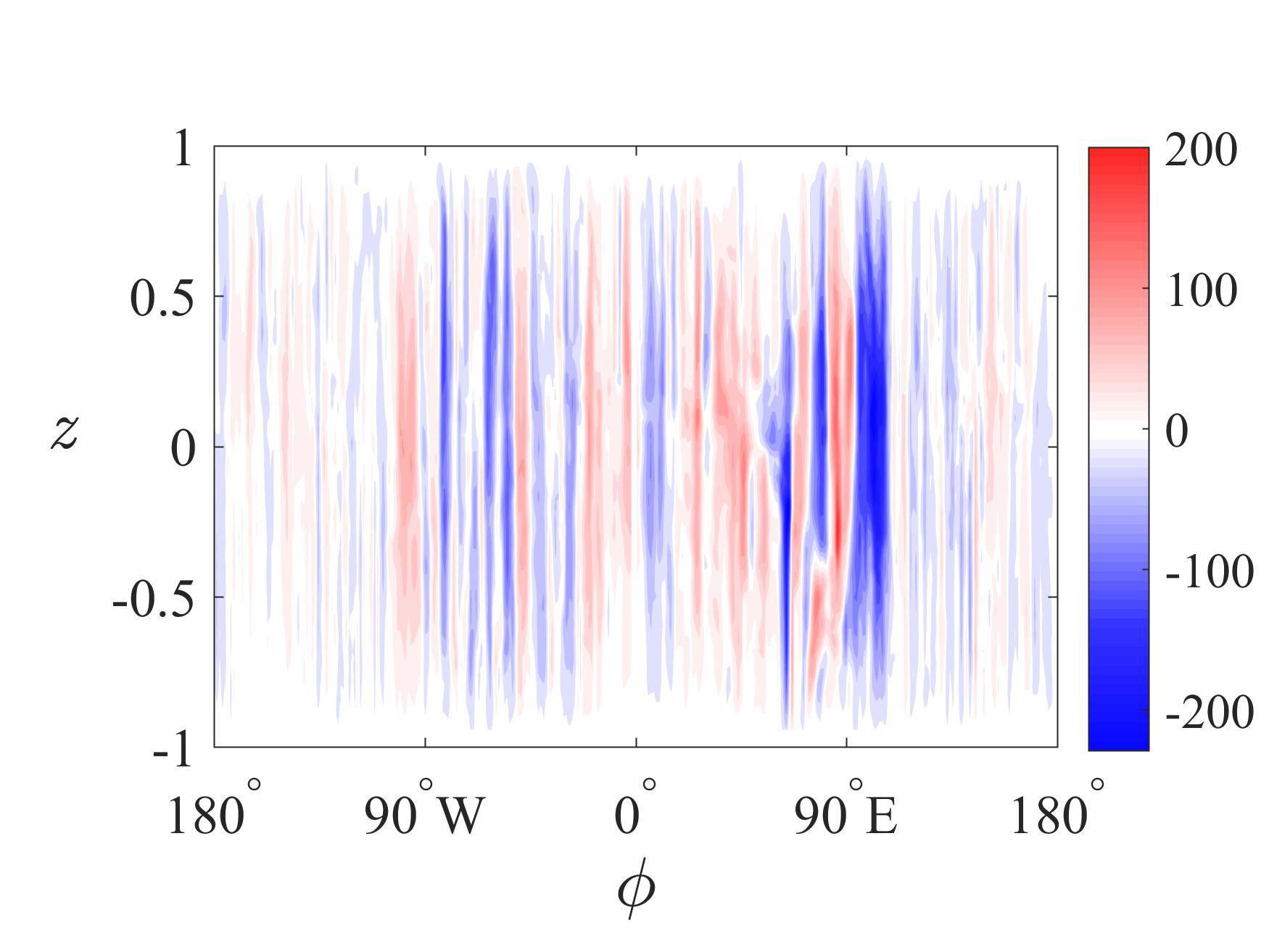}}\\
		\caption{Radial velocity ($u_{r}$) 
		at the equator (a--c) and a cylindrical surface of 
		radius $s=0.8r_{o}$ (d--f) for varying $Ra^C$ in dynamo
		simulations.
			(a) and (d) $Ra^C=1 \times 10^9$,
			 (b) and (e) $Ra^C=1.5 \times 10^9$,
			  (c) and (f) $Ra^C=2 \times 10^9$. 
			  A fixed CMB heterogeneity of $q^{*}=7$ 
			  is used in all simulations. 
			  The common parameters in the simulations
			   are $Ra^T=5.75 \times 10^6 $, $Pr=0.1$, $Sc=1$, $Pm=1$,
			   and  $E= 2 \times 10^{-6}$. 
			   The plots are averaged over 10 turnover times.}
		\label{urplots}
		
	\end{figure}

	The effect of inhomogeneous outer core convection on the freezing rate 
	of the inner core is understood by plotting 
	the heat flux distribution on the ICB in the dynamo (figure \ref{hfplots}). Here, 
	positive values indicate radially outward heat flux 
	while negative values give radially inward heat flux. The line plots in 
figure \ref{hfplots} (b,d,f) show the variation of ICB heat flux ($Q_{ic}$)
	at the equator (blue line) and $30 ^\circ$S (red line). The
variation in heat flux is largely confined to the region between these latitudes, 
and closely follows
the pattern of convection in figure \ref{urplots}. The highest
positive heat flux occurs approximately where the coherent downwelling impinges
on the ICB. The regions of positive heat flux are separated by troughs
of negative (inward) heat flux at the equator beneath Africa and the Pacific.
When equally coherent downwellings exist beneath Asia and America
(figure \ref{urplots} (b)), the peak positive heat flux on
the ICB equator is nearly equal in both hemispheres
 but the negative
 heat flux is more pronounced beneath the Pacific than 
 Africa (figure \ref{hfplots} (d)). The core flow regime 
 with a coherent downwelling beneath Asia and time-varying
 convection beneath America (figure \ref{urplots} (c))
produces peak positive ICB heat flux
	between $80^\circ$E and $100^\circ$E. The heat
flux pattern may be considered to be the superposition of a hemispherical
($m=1$) variation on an underlying, predominantly two-fold
($m=2$) variation. 
 
In all flow regimes, the 
 greatest variation in ICB heat flux occurs 
	in the equatorial region. 
	Inward heat flux is not present beyond $30^\circ$S and $30^\circ$N,
	and the azimuthal variation at higher latitudes is much
	smaller than that at low latitudes. As the regime with
quasi-stationary convection beneath Asia and time-varying
 convection beneath America is relevant to present-day Earth, it
is examined in detail below.

	\begin{figure}[!htbp]
		\centering
		\subfloat[\label{fig:hfq7a}]{
			\includegraphics[width=0.45\textwidth, keepaspectratio]{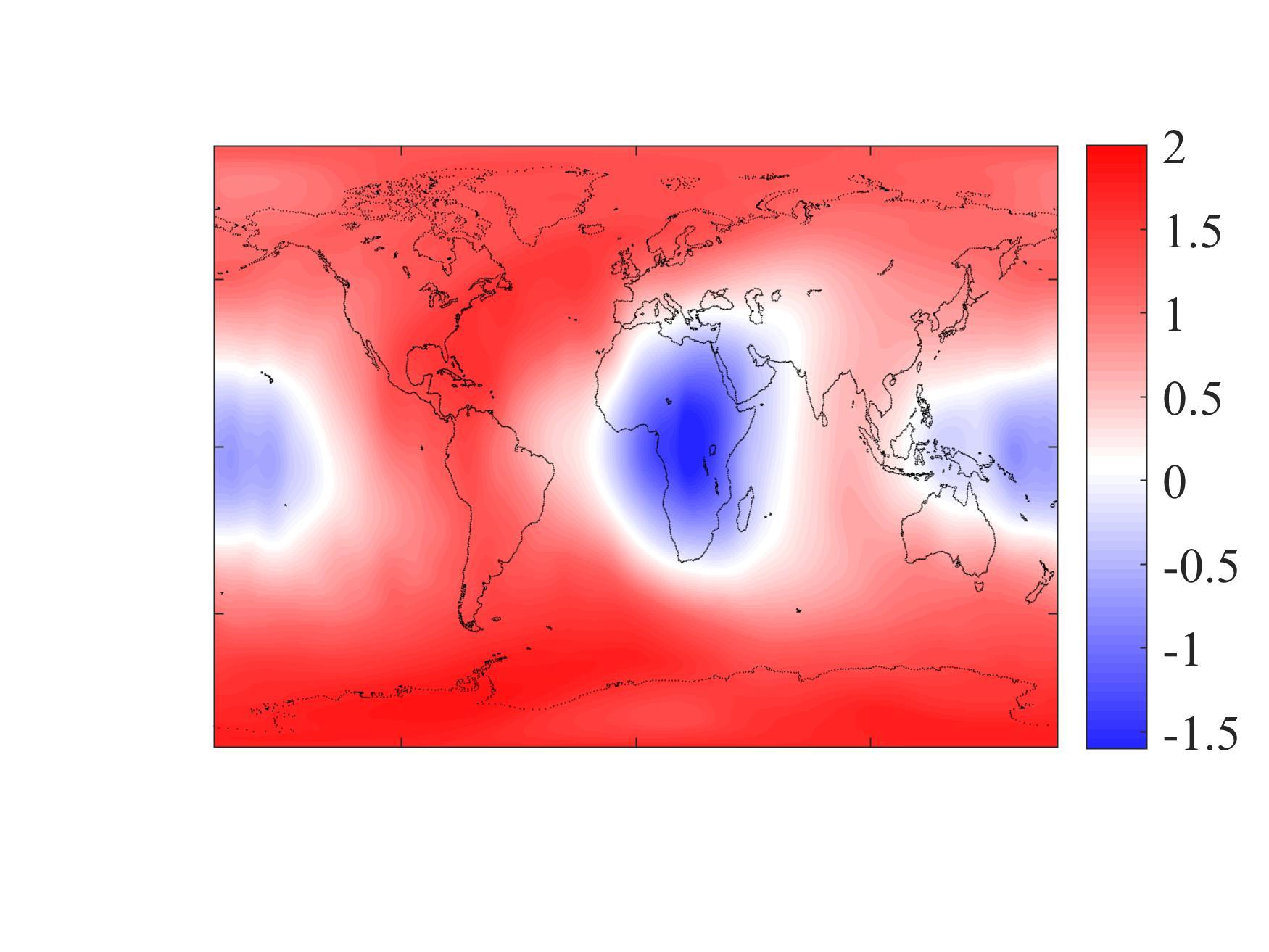}}
		\subfloat[\label{fig:hfq7a_lat}]{
			\includegraphics[width=0.45\textwidth, keepaspectratio]{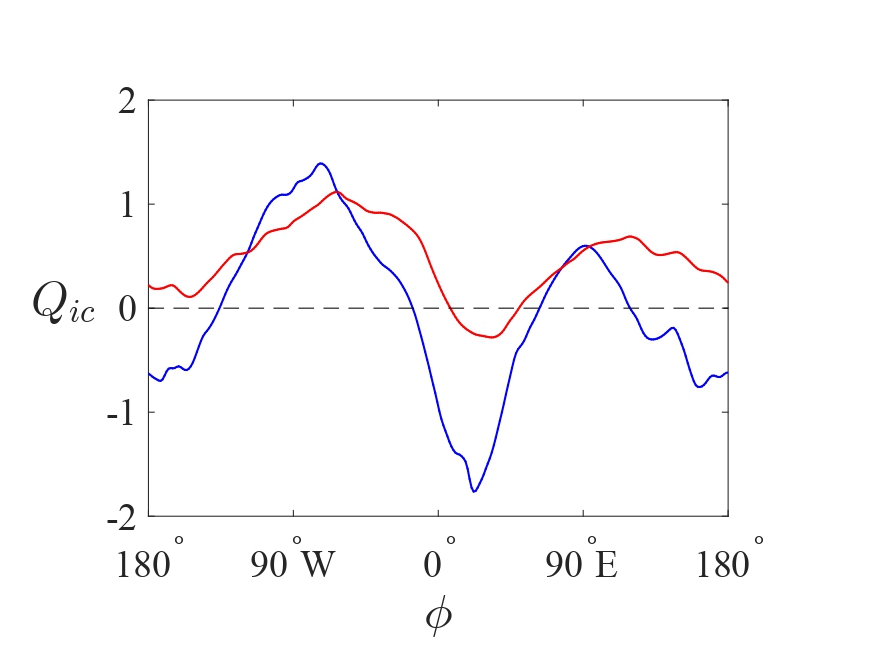}}\\
		\subfloat[\label{fig:hfq7b}]{
			\includegraphics[width=0.45\textwidth, keepaspectratio]{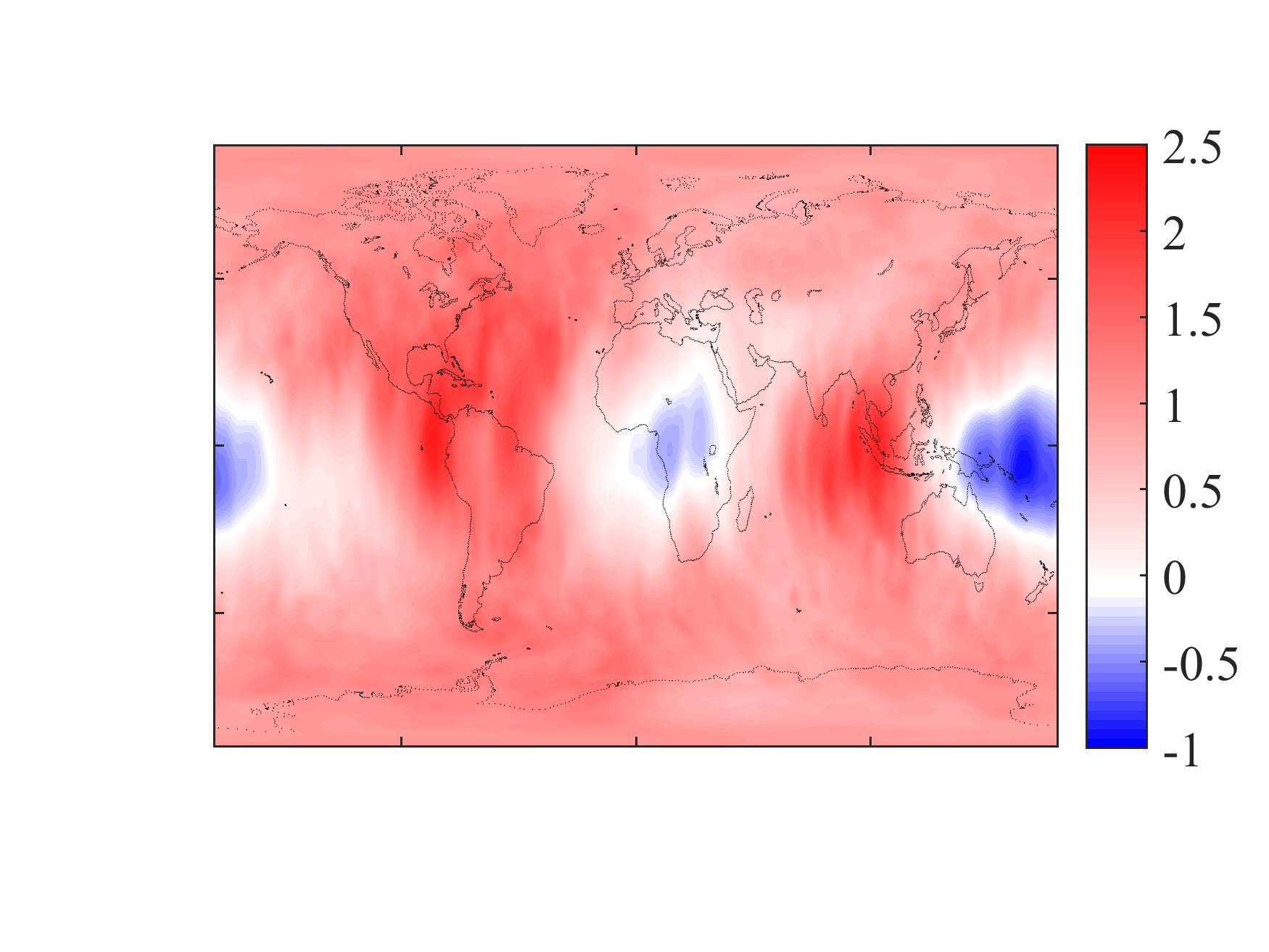}}
		\subfloat[\label{fig:hfq7b_lat}]{
			\includegraphics[width=0.45\textwidth, keepaspectratio]{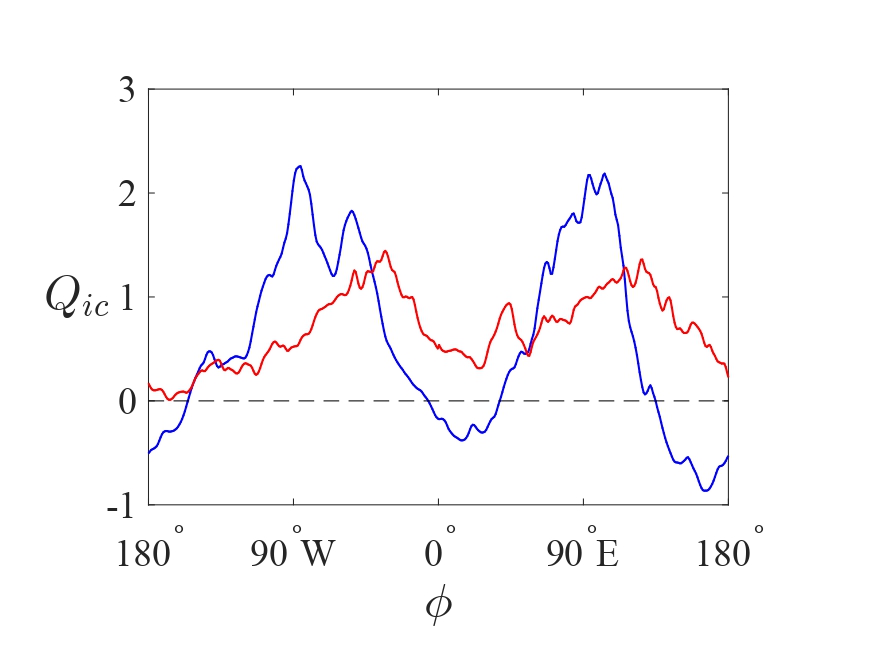}}\\
		\subfloat[\label{fig:hfq7c}]{
			\includegraphics[width=0.45\textwidth, keepaspectratio]{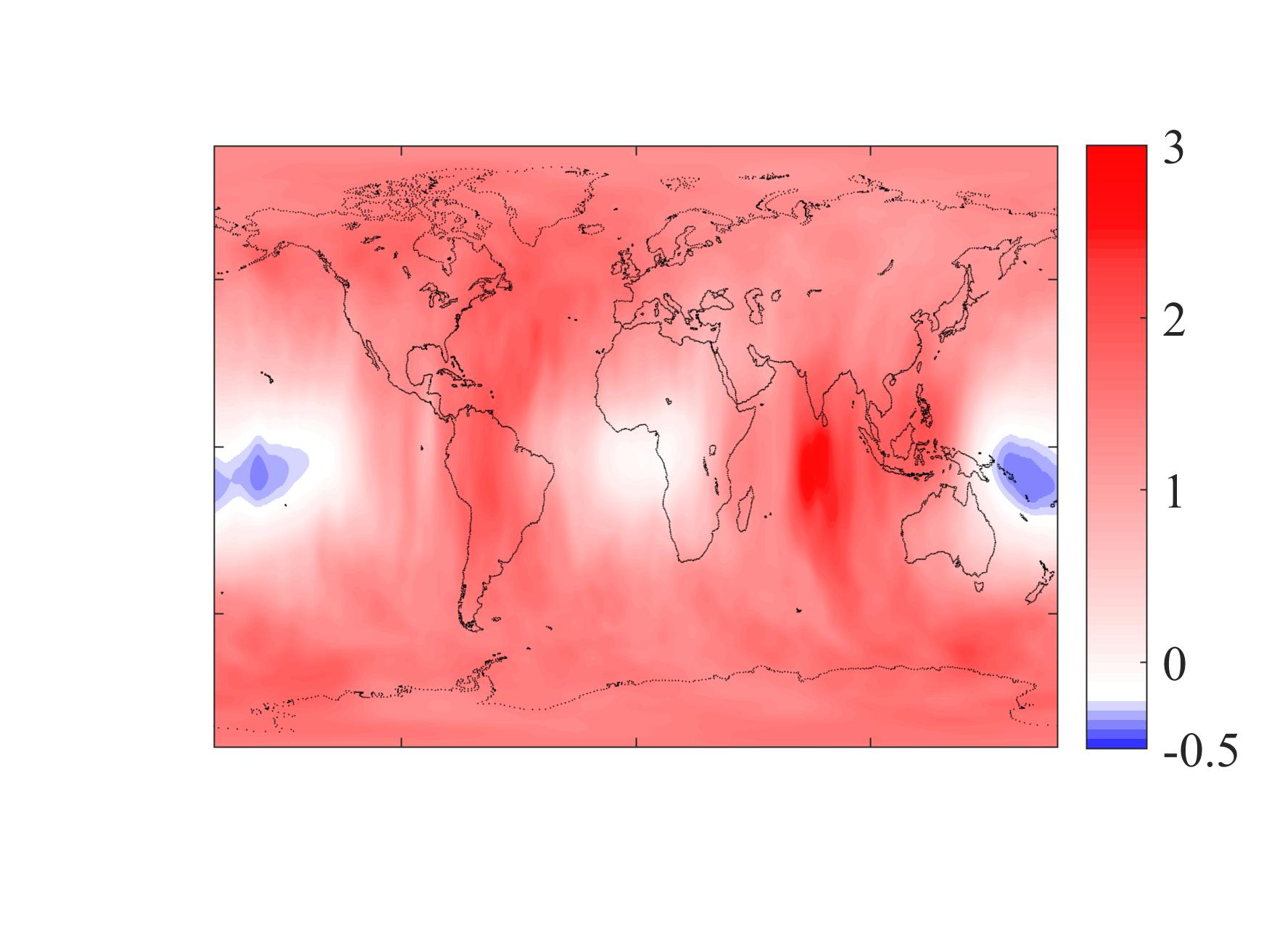}}				    
		\subfloat[\label{fig:hfq7c_lat}]{
			\includegraphics[width=0.45\textwidth, keepaspectratio]{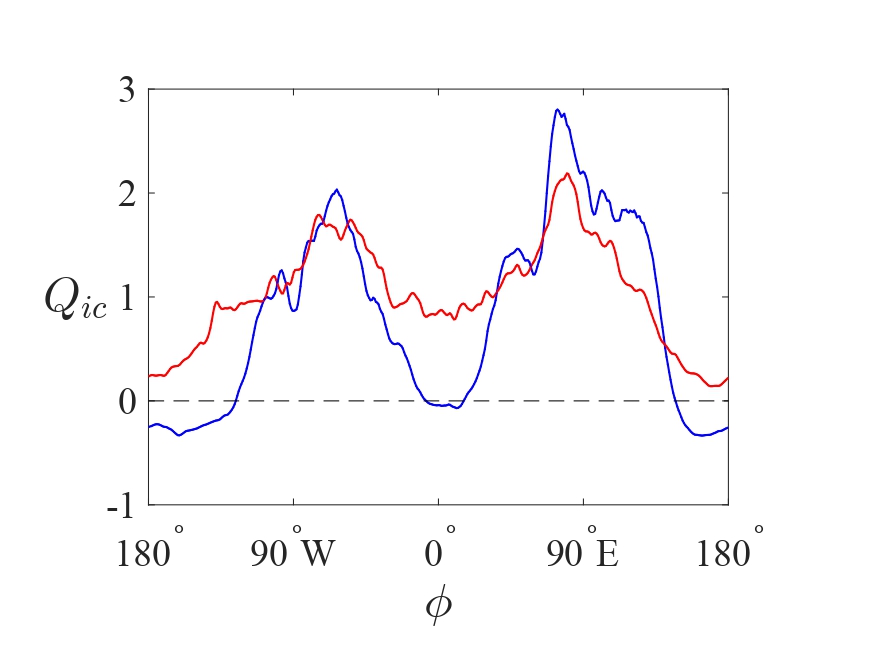}}\\
		\caption{ICB heat flux distribution (left panels)
		 and longitudinal variation of the ICB heat flux (right
		 panels) at the equator (blue line) and  at $30 ^\circ$S (red line), 
		for varying 
		$Ra^C$ at a fixed CMB heterogeneity $q^*=7$ in dynamo simulations. 
		(a) and (b) $Ra^C=1 \times 10^9$, 
		 (c) and (d) $Ra^C=1.5 \times 10^9$, 
		 (e) and (f) $Ra^C=2 \times 10^9$. 
		 The common parameters in the simulations 
		 are $Ra^T=5.75 \times 10^7$, $Pr=0.1$, $Sc=1$, 
		 $Pm=1$, and $E=2 \times 10^{-6}$. The dashed horizontal lines correspond to zero heat flux. The plots are averaged over 10 turnover times.}
		\label{hfplots}
	\end{figure}

	Figure \ref{fig:Q_qcomp} (a) shows the ICB heat flux distribution
in dynamos driven by one-component
convection using basal heating and a uniform volumetric heat sink (see
Appendix A for the model description). 
This buoyancy profile is known to promote boundary effects on the convection
even in strong forcing \citep{sreenivasan2008}. Here, the CMB anomalies do not
reach the ICB even at $q^*=10$, in agreement with strongly forced 
basally heated models \citep{davies2019}. In contrast, the dynamos
driven by two-component convection consistently show the mantle-induced
heterogeneity of the inner core (figure \ref{fig:Q_qcomp} b--d).
The peak ICB heat flux 
	for most simulations is located
	within $80^\circ$E--$100^\circ$E, which is in good agreement with the location
	of peak P-wave velocities in
	\cite{cormier2013}. Regions of higher
	positive heat flux freeze faster, leading to higher
	isotropic P-wave velocities.  \cite{cormier2013} 
	also obtain a local peak beneath
	the Atlantic, which corresponds to the location
	of preferred convection in the Western hemisphere. 
	In general, there are
	two regions of high positive heat flux beneath Asia
	and America and two regions of low or negative
	heat flux beneath Africa and the Pacific.
	
	Simulations with $q^*$ of $O(1)$ and
	$O(10)$ present marked differences in the ICB heat
flux pattern. 
	For $q^*=1$ and $q^*=2$ 
	(figure \ref{fig:Q_qcomp} (b)), 
difference in positive heat flux between the two hemispheres is small; furthermore,
the heat flux does not reverse in sign. For $q^*$ of $O(10)$, the positive
heat flux beneath Asia is systematically higher than that beneath
America; furthermore, the reversal of heat flux is observed dominantly
	beneath the Pacific
	(figure \ref{fig:Q_qcomp} (c)).
	The increased negative heat flux under the Pacific 
	plays a significant role in the east--west dichotomy.
	The region 
	beneath Africa experiences only a marginally 
	negative heat flux although its magnitude might
increase at still higher $q^*$. 
%

	\begin{figure}[!htbp]
		\centering
		\subfloat[]{\includegraphics[width=0.4\textwidth, keepaspectratio]{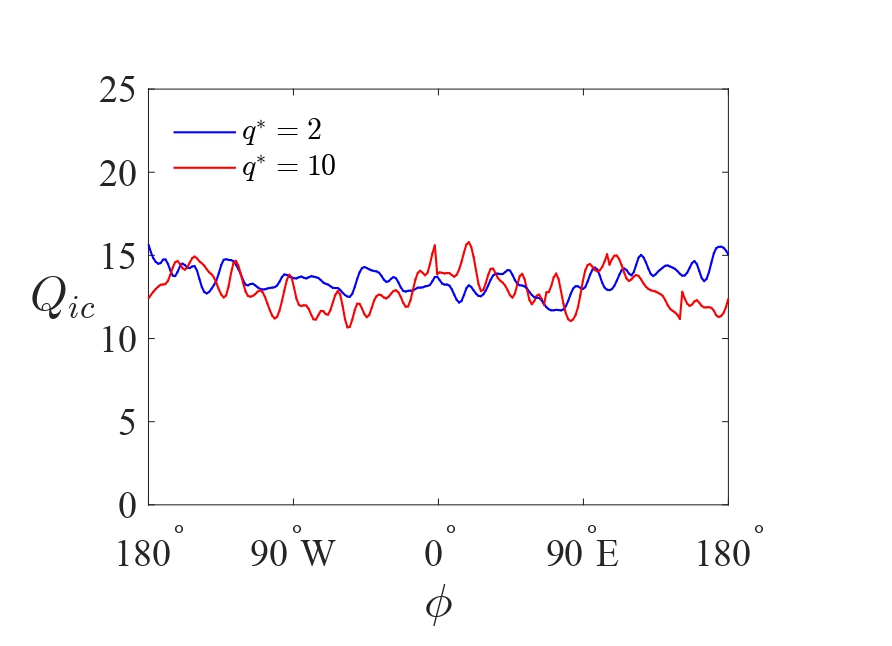}}
		\subfloat[]{
			\includegraphics[width=0.4\textwidth, keepaspectratio]{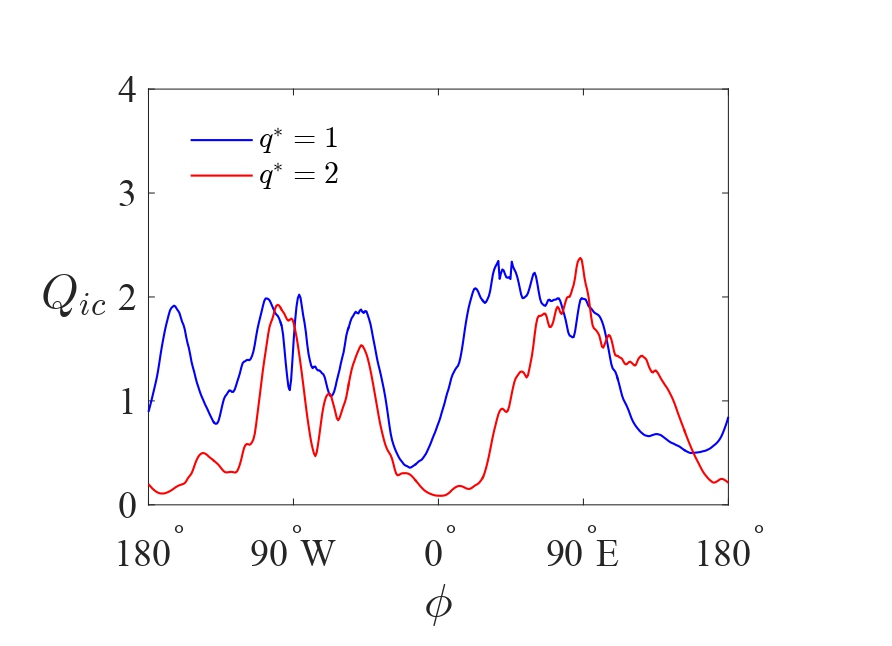}} \\
	\vspace{-0.6cm}
		\subfloat[]{
			\includegraphics[width=0.4\textwidth, keepaspectratio]{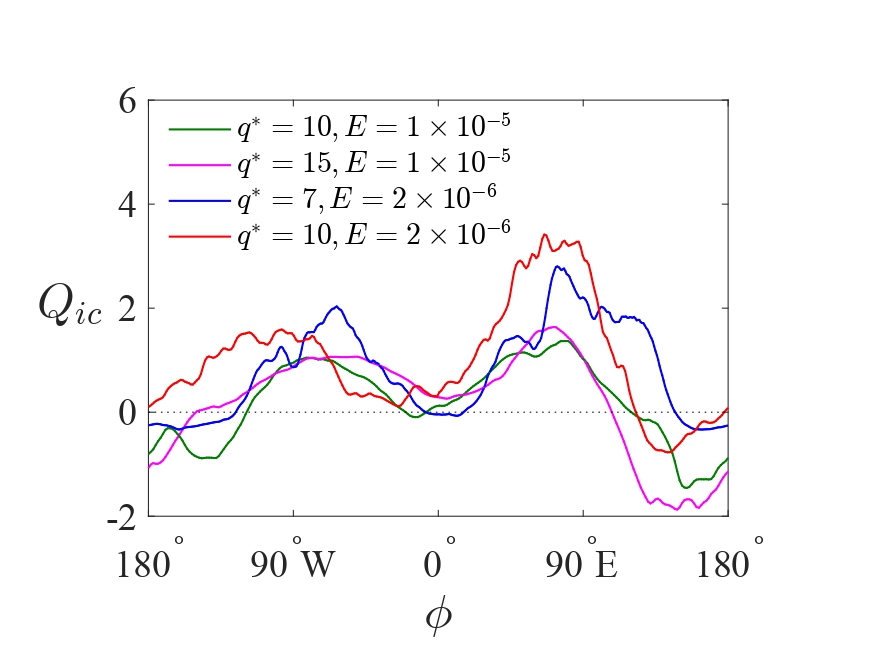}}
		\subfloat[]{
			\includegraphics[width=0.4\textwidth, keepaspectratio]{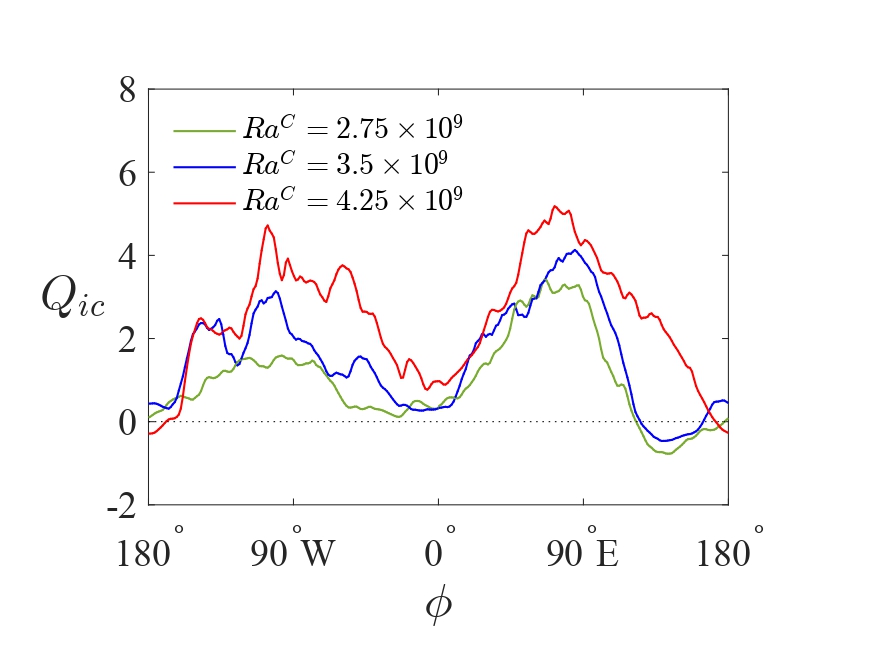}}
			\caption{(a): Heat flux measured at the 
		equator of the ICB ($Q_{ic}$) for two values of $q^{*}$ in the dynamos driven by
one-component convection with parameters
		$Ra=2 \times 10^9$, $Pr=1, Pm=1, 
		E=2 \times 10^{-6}$ (Appendix A). 
		(b)--(d) Heat flux measured at the equator of 
		the ICB  
		in the two-component convective dynamos for different 
		values of $q^*$ in the regime of coherent 
		convection beneath Asia and time-varying
		 convection 
		beneath America. The parameters not given in the
		legends are 
		(b) blue - $Ra^T=7.2 \times 10^6$, $Ra^C=2 
		\times 10^9$, $E=2 \times 10^{-6}$;
		red - $Ra^T= 6.75 \times 10^6$, $Ra^C=2 
		\times 10^9$, $E=2 \times 10^{-6}$, 
		(c)  blue - $Ra^T=5.75 \times 10^6$, $Ra^C=2 
		\times 10^9$; red - $Ra^T= 4.3 \times 10^6$, $Ra^C=2.75 
		\times 10^9$; green - $Ra^T=1.12 \times 10^6$, $Ra^C=0.9 
		\times 10^9$; magenta - $Ra^T=0.9 \times 10^6$, $Ra^C=1.05 
		\times 10^9$, (d) $Ra^T= 4.3 \times 10^6$, $E=2 \times
		10^{-6}$, $q^*=10$ for all three runs. 
		The thermal Rayleigh number 
		$Ra^T$ is kept at its critical value for nonmagnetic
		convection in all simulations.		
		The common parameters are $Pr=0.1$, $Sc=1$ and $Pm=1$.
		 The horizontal dotted black 
		lines in (c) and (d) show zero heat flux. The plots are averaged 
		for 10 turnover times.  }		
		\label{fig:Q_qcomp}
		\end{figure}

	The effect
	of progressively increasing $Ra^C$ at a given $q^*$ on the ICB heat flux
	is given in figure \ref{fig:Q_qcomp} (d).
	Increasing $Ra^C$  causes the positive heat flux
	to increase overall, resulting in a higher average heat flux. 
	The negative heat flux beneath Africa and the Pacific is reduced as a result. 
	The difference in average 
	heat flux between the eastern and western
	hemispheres reduces as $Ra^C$ is increased.

	\subsection{East-West differences in freezing rate and seismic
		velocity} \label{seismic}
	

	The differences in heat flux between the eastern and 
	western hemispheres manifest as variations in 
	freezing rates and P-wave speeds within the upper 
100 km of the inner core \citep{niu2001}. 
	To estimate these variations, we 
	focus on the ICB region between
	$30^\circ$S and $30^\circ$N.  
	These latitudinal bounds are chosen because they encompass
	the regions of enhanced and reversed heat flux
	observed at the ICB. The high latitudes exhibit
	little lateral
	variation and it is unlikely
	that high $q^*$ accompanied by high $Ra^C$
	would widen the range of latitudes
	of ICB heterogeneity. The choice
	of the latitudinal bands 
	is also justified by seismic studies
	 \citep[e.g.][]{ohtaki2012}, where
	the highest compressional velocity is confined
	to equatorial East Asia. This is broadly
	consistent with the location of the peak
	heat flux in our dynamo simulations.
	
To obtain the average density at the top of the inner core
in the eastern and western hemispheres, $\rho_e$ and
$\rho_w$, one must first
estimate the mass fractions of solid and liquid iron,
$x_s$ and $x_l$ respectively, in either
hemisphere, which is in turn determined by the ICB heat flux measured
in the dynamo. The solid mass fraction in the east, $x_{s,e}$,
is first estimated as follows. The solid mass fraction 
at any point is taken to be unity
for heat flux 
$Q \ge \overline{Q}_{ic}$, where $\overline{Q}_{ic}$ is the
surface-averaged heat 
flux between $30^\circ$S and $30^\circ$N. For $Q \le 0$, the solid
mass fraction is zero since inward heat flux at any point
is assumed to cause melting. For $0 < Q < \overline{Q}_{ic}$,
the solid mass fraction varies linearly with $Q/\overline{Q}_{ic}$.
The mean solid mass fraction in the eastern hemisphere is subsequently 
      determined by averaging the 
      mass fraction over all points between $30^\circ$S and $30^\circ$N.
The hemispherical solid mass fractions are related to the
respective mean ICB heat fluxes by
	\begin{equation} \label{mfrac}
		\dfrac{x_{s,e}}{x_{s,w}}=\dfrac{E_{avg}}{W_{avg}},
	\end{equation}
which gives the solid mass fraction in the
west, $x_{s,w}$. 

The density of the 
	solid phase of iron over the ICB, $\rho_{s}=12704$ kg m$^{-3}$,  
	is taken from the reference model AK135 \citep{kennett1995}.
	 The density of liquid iron is given the value
	  $\rho_{l}= \rho_{s}
	- \Delta$, where $\Delta=240$ $\mathrm{kg\hspace{1pt}m^{-3}}$
\citep{alfe2002a}.
While it has been argued that the density jump at the ICB 
	might be near 300 $\mathrm{kg\hspace{1pt}m^{-3}}$ 
	\citep{irving2018,robson2019}, the value
	taken in this study
	lies in the range 200--300 $\mathrm{kg\hspace{1pt}m^{-3}}$
	\citep{koper2004,tkalvcic2009}, and is reasonable if
	one assumes that the change in density is mainly
	due to solidification rather than differences in 
	composition between the solid and liquid phases. 
	The mean density for either hemisphere
	is calculated from the densities of the two pure
	phases by the method of \lq Reuss averaging'
	\citep[p.~125]{anderson1989},
	\begin{equation}
		\rho=\dfrac{1}{\dfrac{x_{s}}{\rho_{s}}+\dfrac{x_{l}}{\rho_{l}}},
	\end{equation}	
where $x_l=1-x_s$. The mean bulk modulus $K$ for either hemisphere
is calculated by the same method, where the bulk moduli of
the solid and liquid phases, $K_s$ and $K_l$, are obtained
from the model AK135.  	
The body wave compressional velocity for either hemisphere
is then given by
	\begin{equation} \label{eq:pwavevel}
		v_p =\sqrt{\frac{K}{\rho}},
	\end{equation}
	where the shear modulus of the mixture  is ignored in the 
	calculation of $v_{p}$ \citep[see][]{gubbins2008}.

	The  difference in seismic velocity 
	between the eastern and western
	hemispheres is given by 
	
	\begin{equation}
	 	\Delta v_{p}= v_{p,e}-v_{p,w}.
	\end{equation}
Using the
	values of $K_s$ and $K_l$ from the reference
	model PREM \citep{dziewonski1981}
	gives a lower value of $\Delta v_{p}$ than that obtained
	using AK135 because the higher value of $K_l$ in PREM
lowers the velocity contrast between the
	solid and liquid phases.
	
   The relative difference in freezing rate between the hemispheres 
    is given by

    	\begin{equation}
    	\Delta r_{f}= (E_{avg}-W_{avg})/\overline Q_{ic}.
    \end{equation}


	
	
%
	
	
	

	
	
	\begin{table}[!htb]
	\centering
	\begin{tabular}{|c|c| c| c| }
		\hline
		Author &  Reference model - $V_{p}$   &  $\Delta V_{p}$  in \% ( in $ \mathrm{m\hspace{1pt}s^{-1}}$) & Boundaries\\
		
		\hline
		\cite{niu2001} & PREM - 11.03 $\mathrm{km\hspace{1pt}s^{-1}}$ & 0.8 (88) & $40^\circ$E, $180^\circ$E\\
		\cite{sun2008} & AK135 - 11.04 $\mathrm{km\hspace{1pt}s^{-1}}$ & 0.5 (55) & $40^\circ$E, $160^\circ$E\\
		\cite{waszek2011} & AK135 - 11.04 $\mathrm{km\hspace{1pt}s^{-1}}$  & 1.5 (165)  & 10$^\circ$--14$^\circ$E, $173^\circ$E\\
		\cite{burdick2019} & AK135 - 11.04 $\mathrm{km\hspace{1pt}s^{-1}}$  & 1.0 (110) & $24^\circ$E, $180^\circ$E\\
		\hline
	\end{tabular}
	
	\begin{tabular}{|c|c|}
		\hline
		Author & F-layer height (km)\\
		\hline
		\cite{souriau1991}  & 150 \\
		\cite{kennett1995}  & 150 \\
		\cite{zou2008}  &   350 \\
		\cite{ohtaki2015}  & 230-380 \\
		\hline
		
	\end{tabular}
	\caption{Top panel: Summary of studies listing the
		difference in average 
		isotropic P-wave velocity 
		between the quasi-eastern
		and quasi-western hemispheres in 
		the outermost region of the inner core. 
		The absolute difference between the eastern 
		and western hemispheres ($\Delta v_p$) is 
		calculated using the hemispherical
		differences in per cent and the reference velocity
		used in each work. The boundaries 
		dividing the quasi-eastern and quasi-western 
		hemispheres are also given.
		Bottom panel: Estimates of the F-layer height proposed by 
		different seismic studies. }
	\label{tab:seismic_references}
\end{table}	

\begin{table}[!htb]
	\resizebox{14.5cm}{!}{
		\centering
		\small
		\begin{tabular}{|c|c|c|c c | c c c|c c| c c| c c |c|c|}
			\hline
			Sr. no.& $E$& $q^*$   &  $Ra^T$ & $Ra^C$    & \multicolumn{3}{|c|}{Heat flux} & \multicolumn{2}{|c|}{Mass fraction} & \multicolumn{2}{|c|}{Density}  & \multicolumn{2}{|c|}{Bulk modulus} & & 
			\\
			&  &  &        & &$E_{avg}$ & $W_{avg}$ & $\overline Q_{ic}$ &$x_{s,e}$ & $x_{s,w}$ & $\rho_{e}$ & $\rho_{w}$ & $K_{e}$ & $K_{w}$ & $\Delta r_{f} (\%)$ & $\Delta v_{p}$  \\
			& & & $(\times 10^6)$  & $(\times 10^9)$     & & & & & & \multicolumn{2}{|c|}{($\mathrm{kg\hspace{1pt}m^{-3}})$}   &  \multicolumn{2}{|c|}{(TPa)}  & & $(\mathrm{m\hspace{1pt}s^{-1}})$\\
			\hline
			$a$& $2 \times 10^{-6}$&1   &  7.2  & 2  &  1.38  & 1.32 & 1.35 & 0.80  & 0.77 & 12655 & 12647 & 1.330 & 1.328 & 3.77  & 4   \\
			&  &  &   &       &  (1.23)  & (1.13) & (1.18) &(0.71) & (0.65) & (12633) & (12620) & (1.325) & (1.315) & (8.42)  & (7)   \\
			\hline
			$b$& & 2    &  6.75 & 2 &  1.14  & 1.07  & 1.10 &0.75 & 0.70 & 12643 & 12630& 1.327  & 1.324  & 6.60 & 7 \\  		
			&   & &  &       &  (0.90)  & (0.77) & (0.83) &(0.63) & (0.54) & (12614) & (12592) & (1.320) & (1.315) & (15.57)  & (12)  \\ 
			\hline 	
			$c\dagger$ & & 7   & 5.75    & 2  &  1.04 & 0.83 & 0.94 & 0.73 &0.56 & 12638& 12603 &1.326  & 1.317& 22.64  & 21\\
			&    &  &   &   &  (1.03) & (0.60) & (0.82)&(0.67) &(0.48) & (12624)& (12557) &(1.323)  & (1.307)& (52.60) & (35)\\
			\hline
			$d$ & &7   & 5.75    & 2.75  &  1.39 & 1.20 & 1.30 & 0.69 &0.60 & 12629& 12606 &1.324  & 1.318& 14.73  & 12\\
			&    & &    &   &  (1.26) & (1.01) & (1.13) &(0.60) &(0.48) & (12607)& (12598) &(1.318)  & (1.312)& (22.03) & (15)\\
			\hline         
			$e^\dagger$& &10    & 4.30  & 2.75 & 1.25 & 1.01 & 1.13 & 0.70 & 0.57  & 12631& 12599 &1.324 & 1.317 & 20.75 &17\\
			&       &   &    &      & (1.21) & (0.75) & (0.98)& (0.58) &( 0.35)  & (12602)& (12549) &(1.318) & (1.305) &( 46.94) & (28)\\
			\hline
			$f$& & 10  & 4.30 & 3.5  &  1.89  & 1.78 & 1.84 & 0.75 & 0.71 & 12643 & 12633& 1.327 & 1.325 & 5.98  & 6 \\
			&     & &      &     &  (1.80)  & (1.60) & (1.75) &(0.65) & (0.58) & (12619) & (12602)& (1.322) & (1.317) & (11.76) & (11) \\
			\hline
			$g$ & &10  & 4.30 & 4.25  & 2.87& 2.75 & 2.81 & 0.83 & 0.80  & 12663 &12654 &1.332& 1.330 & 4.27 & 4\\
			&    & &    &     & (2.60) & (2.30) &(2.45) & (0.74) & (0.65)  & (12641) &(12620) &(1.327)& (1.322) & (12.24)& (11)\\
			
			\hline
			\hline
			$h$ & $1\times 10^{-5}$ & 10  & 1.12 & 0.9  & 0.56 & 0.46  & 0.51& 0.70 & 0.58  & 12631 &12601 &1.324& 1.317 & 19.61& 16\\
			&    & &     &    & (0.26) & (0.15) & (0.20) &(0.50) & (0.29)  & (12585) &(12534) &(1.313)& (1.301) & (52.31)& (28)\\
			$i$ & & 15  & 0.87 & 1.05  & 0.39 & 0.33 & 0.36 & 0.67 & 0.36  & 12624 &12599 &1.322& 1.316 &16.67& 14\\
			&  & &  &               & (0.09) & (0.06) & (0.075) & (0.61) & (0.41)  & (12609) &(12560) &(1.319)& (1.307) &(40.00)& (26)\\
			\hline
	\end{tabular}}
		\caption{Differences in average freezing rate $\Delta r_{f}$
		(per cent)   and P-wave velocity difference
		$\Delta v_p$
		(m s$^{-1}$) 
			between the eastern and western
			hemispheres evaluated between $30^\circ$S
			and $30^\circ$N
			of the ICB. 
			The values in the brackets
			 are those calculated for the equator.
			The regime of stable convection beneath Asia 
			and time-varying convection beneath the Atlantic is 
			considered in all runs. Here,
			$x_{s,e}$ and $x_{s,w}$ are the solid mass 
			fractions in the eastern and western 
			hemispheres of the inner core, 
			$\rho_e$ and $\rho_w$ are
			the respective mean densities and
			$K_{e}$ and $K_{w}$ are the respective mean
			bulk moduli.  
			The superscript $\dagger$ denotes the parameters 
			for which the reversal of heat flux and  
			$\Delta r_{f}$, 
			$\Delta v_{p}$ 
			are the highest among the runs performed for that 
			$q^*$. The dynamo 
			parameters are $Pr=0.1$, $Sc=1$, $Pm=1$, 
			$E=2 \times 10^{-6}$ for (a)--(g) and 
			$Pr=0.1$, $Sc=1$, $Pm=1$, 
			$E=1 \times 10^{-5}$ for (h)--(i). }
		\label{tab:seismic_sim}
		
	\end{table}

    Previous studies that map the isotropic P-wave velocity
	at the top of the inner core report a difference of 
	0.5-1.5\% between the two hemispheres, 
	which translates to absolute differences
	of $\approx$ 50--160 $\mathrm{m\hspace{1pt}s^{-1}}$
	(table \ref{tab:seismic_references}). 
	In these studies, the boundaries 
	demarcating the eastern and western hemispheres 
	are not located at $0^\circ$ 
	and  $180^\circ$; rather, \lq quasi-eastern' and 
	\lq quasi-western' hemispheres are considered 
	\citep[e.g.][]{tanaka1997}
	based on the magnitudes of the P-wave velocity
	relative to its reference value prescribed by
	PREM \citep{dziewonski1981}
	or AK135 \citep{kennett1995}.  
	Table \ref{tab:seismic_sim} gives the east--west differences 
	in the freezing rate and
	isotropic P-wave velocity at the ICB
	using the hemispherical heat fluxes in
	the dynamo simulations. 	
	The differences 
	($\Delta r_{f}$ in per cent and $\Delta v_{p}$ in m s$^{-1}$) 
	are estimated for the conventional hemispherical 
	boundaries at [$0^\circ$, $180^\circ$] and
	latitudes [$30^\circ$S, $30^\circ$N]. 
	For $q^*$ of $O(10)$, the values
	of $\Delta v_{p}$ at the equator, given within brackets,
	do not reach the the lower bound
	of the observed variation (table 
	\ref{tab:seismic_references}) but are of the
	same order of magnitude.
	Since the dynamo model has no reference
	seismic velocity, it would not be logical to
	estimate the east--west differences for skewed boundaries 
	between the hemispheres.
	 Nevertheless, when
	the averaging in \eqref{mfrac} is performed
	over a truncated eastern hemisphere,
	the hemispherical differences increase 
	because part of the region with negative heat 
	flux in the Pacific at $q^*$ of $O(10)$
	now lies in the
	quasi-western hemisphere. Consequently, this 
	lowers $W_{avg}$ relative to $E_{avg}$, 
	resulting in an increased contrast between 
	the two hemispheres. For example, a variation of
	$\Delta v_p \approx 50$ m s$^{-1}$ is obtained
	for $q^* =$10 with the quasi-eastern boundaries
	[$40^\circ$E, $160^\circ$E]. This is near
	the lower bound of the observed variation. 

	The values of $\Delta r_{f}$ and 
	$\Delta v_{p}$ at $q^*$ of $O(1)$ and 
	$O(10)$ suggest that 
	the CMB must have a heterogeneity  
	$q^*$ of at least $O(10)$.

		\subsection{Height of the 
melt-induced stratified layer at the base of the outer core}
		
		   	A stably stratified layer 
		   	at the base of the Earth's 
		   core 
\citep{souriau1991,zou2008,ohtaki2015}, also known as the F-layer,  
		   can be explained by localized melting 
		   of the inner core \citep{gubbins2008,gubbins2011}.
Here, 
outer core convection causes heat flow
into the inner core primarily beneath the Pacific and to a lesser extent
beneath Africa, which in turn produces the melt that feeds the layer.
The shallow water approximation for an incompressible Boussinesq
fluid \citep{deguen2014f} subject to buoyant
forcing and magnetic damping provides a useful model for the F-layer.

		The conservation
		of mass in a layer where the density decreases linearly
from bottom to top is given by
		
		\begin{equation}	\label{eq:massbal}	
			\dfrac{\partial h}{\partial t} + \nabla \cdot 
			({\bm u} h)=2 w,
		\end{equation}
		
		where ${\bm u}$ is the horizontal velocity averaged
			over the layer height $h$,
			$w$ is the volume flux per unit area into
		the layer from the inner core and the mean
		variation in density across the layer is half the
		difference in density between the top and bottom. 
		The ICB heat flux variation in
		figure \ref{fig:Q_qcomp} (c) suggests that $w$ would
		take the approximate form
		$w_0 \cos 2\phi$, where the peak volume flux density $w_0
		\sim \kappa_T/h_0$ from the steady-state heat
		equation, $\phi$ is the longitudinal
		separation from the point of peak inward heat flux
		and $h_0$ is
		the peak layer height.

		 The equation
		of motion in the layer is given by
		
		\begin{equation}		\label{eq:forcebal2}		
			\rho \left( \dfrac{\partial \bm u}{\partial t}
			+ 2 \bm{\varOmega} \times \bm{u} \right)  = 
			- {g \delta \rho} \nabla h + 
{\bm j} \times {\bm B},
		\end{equation}		
		
		where $\varOmega$ is
		the angular velocity of rotation, $j$ is the 
		induced electric current density, 
		$\rho$ is the mean density of the layer and
		$\delta \rho$ is the the variation of density across
		the layer. Both nonlinear inertial and viscous
		forces are small relative to the Coriolis
		force in the rapidly rotating core; 
		that said, since the east--west variation is thought
		to be maximum near the equator (figure \ref{hfplots}),
		 the analysis is confined to the equator where the Coriolis
		force does not enter the force balance. 
		In the steady
		state, the lateral buoyancy is balanced
		by the Lorentz force in the layer. Given that
		density disturbances of length scale $\sim 10$ km,
		of magnetic Reynolds number $Rm \sim$ 1--10, can
		generate the field \citep{jfm21},
		the estimation of the Lorentz force in the low-$Rm$
		approximation is appropriate \citep{94mofflop}. 
		Since the $s$ and $\phi$ (in cylindrical
		coordinates $s,\phi,z$) magnetic
		fields are zero at the equator,
		the ambient field $B$ is approximated
		by a time and azimuthally averaged $z$ field at the
		equator in the saturated dynamo. We
		obtain
		\citep{sreeni2000}
		\begin{equation}
	{\bm j} \times {\bm B} \sim \sigma B_z^2  
	\ell^* \bm{u},
			\label{lzf}
		\end{equation}		
where the curl of Ohm's law is used to estimate the electric
current, $\sigma$ is the electrical
conductivity and
$\ell^*= \ell_\perp/\ell_\parallel$ is the ratio of 
length scales perpendicular and parallel to the ambient field direction. 	

In the steady state, equations
(\ref{eq:massbal}) and (\ref{eq:forcebal2}) may now be combined to
obtain the following dimensionless
equation for the layer height at the equator:
		
		\begin{equation}		\label{eq:hnondim}
	\nabla^2 {h^*}^2 =
	-\dfrac{ 4 \ell^* {B_z^*}^2 }{  h_0^* \, Ra_H}
			\, \cos 2\phi .			
		\end{equation}
where $h^*$ and $B_z^*$ denote the
layer height and field intensity scaled as in the dynamo model
and the modified horizontal Rayleigh number $Ra_H = g \alpha \Delta Q_H  L^2/
2 \varOmega \kappa_{T}$ is based on the average difference in
equatorial ICB heat flux between Asia, 
where the model suggests predominant freezing, and the
Pacific, where the model suggests predominant melting. The
Asian heat flux is evaluated over [30$^\circ$E, 120$^\circ$E]
and the Pacific heat flux over [120$^\circ$E, 150$^\circ$W],
consistent with the longitudinal 
separation of approximately 90$^\circ$ between the peak
and trough of the equatorial heat flux at high $q^*$
(figure \ref{fig:Q_qcomp}(c)).

Equation \eqref{eq:hnondim} is solved subject to the conditions
$h^2=h_{0}^2$ at $\phi=0$ and $h^2=0$ at $\phi=\pi/2$, indicating
that the layer height goes to zero at the point of peak
freezing. The F-layer height is then given by

		\begin{equation}		
		\label{eq:hnondimfinal}
			{h^*}^2 \sim  
			{h^*_0}^2+ \dfrac{{B_z^*}^2 \ell^* {r_{i}^*}^2 
			(\cos 2\phi -1)}{Ra_H h_0^*},			
		\end{equation}
		where 		
		\begin{equation} 	\label{eq:honondim}
			h_0^* \sim
			\left(\dfrac{2{r^*_i}^2 \ell^* {B_z^*}^2 }
			{Ra_H}\right)^{1/3}
		\end{equation}
		gives the maximum height of the layer. For flow length
		scales $\ell_\perp \sim 10$ km within the F-layer where
		$\ell_\parallel \sim 100$ km, the ratio
		$\ell^* \sim 0.1$.
		
		The mean scaled height $\overline {h^*}$ given in
		table \ref{tab:table_flayer} 
		is obtained from the distribution of $h^*$ 
		over the surface of the inner core.
		A layer height $\overline {h}$
		$\approx$ 200 km, obtained for $q^*=10$,
		lies within the 
		range of 150--400 km proposed by earlier
		studies (see table \ref{tab:seismic_references}). 
		On the other hand, the
		mean height of a viscously damped layer
		\citep{deguen2014f} is
		$\approx$ 100 km.
		Although the present
		 analysis is based on orders of magnitude, it
		 supports the existence of a melt-induced and
		 magnetically damped layer at the base of
		 the outer core.

\begin{table}[!htb]
	\resizebox{15cm}{!}{
		\centering			
		\begin{tabular}{|c|c|c|c c | c |c |c| c| c|}
			\hline
			S. no. & $E$ & $q^*$    & $Ra^T$ & $Ra^C$ & $\Delta Q_{H}$  
			&$ Ra_{H}$ &${B_z^*}^2$  
			&  $\overline {h^*}$ &  $\overline{h}$ (km)  \\
			& &  & $(\times 10^6)$    &  $(\times 10^9)$  & & & & &   \\
			
			\hline		
			$a$ &$2 \times 10^{-6}$ & 7   & 5.75    & 2   & 1.67  & 12.47  &0.42  &  0.082 &185
			\\
			$b$ & &7   & 5.75    & 2.75   & 1.94  & 14.48  &0.52  &  0.083& 188\\
			$c$ & &10    & 4.30  & 2.75   & 2.23   &12.45  &0.50  & 0.086&194  \\
			$d$& &10    & 4.30  & 3.5   & 2.91   &16.25  &0.66  & 0.087 &196 \\
			$e$& &10    & 4.30  & 4.25   & 2.48  &13.84  &0.82  & 0.098& 221\\
			$f$& $1\times 10^{-5}$&10    & 1.12  & 0.9   & 1.51  &10.78  &0.58  & 0.095& 215\\
			$g$& &15    & 0.87  & 1.05   & 2.40  &13.56  &0.64  & 0.091& 205\\
			\hline
	\end{tabular}}		
				\caption{Average 
				height of the F-layer scaled by the depth
				of the outer core
					($\overline {h^*}$) in the dynamo
					 simulations 
					for varying $q^*$ in the regime of 
					coherent 
					convection beneath Asia and
					time-varying convection 
					beneath America.
					The common parameters 
					in the simulations are $Pr=0.1$, $Sc=1$
					and $Pm=1$. The 
					 thermal Rayleigh number
					 $Ra^T$ is 
					set to its critical value.
	The modified horizontal Rayleigh number $Ra_H$ is based on $\Delta Q_H$,
	the difference in measured average heat flux between
	the Asian and Pacific sectors of the ICB at the equator.}
				\label{tab:table_flayer}
			\end{table}

	\subsection{High-latitude magnetic flux at the CMB}
	
	For the dynamo with $Ra^T =4.3 \times 10^6$ ($\approx 
	Ra^T_{cr}$), $Ra^C =2.75 \times 10^9$
	($\approx 180 \times Ra^C_{cr}$) and $q^{*}=10$,  
	a long-lived pair of high-latitude flux lobes exist
	in the East and highly time-varying lobes are found
	in the West. While the 
	coherent downwelling beneath Asia concentrates
	quasi-stationary magnetic flux, the time-varying 
	convection beneath America produces
	flux patches which are relatively unstable. 
	The time-averaged $B_{r}$ at the CMB
	is shown in figure \ref{fig:Br_plots}(b). 
	The pair of lobes in the Western hemisphere 
	appear fragmented on time average due to their variable 
	locations at different times. There are also times when convection
	beneath the Atlantic is stable and gives an
	approximately $m=2$ pattern of
	magnetic flux lobes.
	The location of the flux lobes is in fair agreement
	with that at the Earth's CMB 
(figure \ref{fig:Br_plots}(a)), where the magnetic field 
is obtained from the CALS10k.2 
	model compiled from archaeomagnetic and paleomagnetic data
	for the past 10000 years \citep{constable2016}.
	The time average is for the period 1590--1990 AD. While
	Earth-like fields are not produced for
	large CMB heterogeneity in
	weakly rotating dynamos with $E \sim 10^{-4}$
	 \citep{driscoll2015}, the rapidly rotating
	dynamos with $E \sim 10^{-6}$ considered here can reproduce
	Earth-like fields even for $q^*$ of $O(10)$. 

	Figures \ref{fig:Br_plots} (c) \& (d) 
	show the longitudinal 
	positions of the peak magnetic flux in the 
	Northern hemisphere lobes. 
	The east--west variability 
	in the dynamo simulation is compared with that of the 
	archeomagnetic field model CALS10k.2 \citep{constable2016} 
	from 4000 BC to 1990 AD and the CHAOS-7.12 model from 1990 to 
	2020 \citep{finlay2020}. The solid lines in figure 
	\ref{fig:Br_plots} (c) show the historical peaks of the magnetic
	flux lobes for the past 6000 years.
In the past 3000 years, the western 
	hemisphere lobes have been more mobile than their eastern
	hemisphere counterparts -- while the Canadian
	flux lobe moved between $50^\circ$W to $100^\circ$W,  the Siberian
	lobe moved between $90^\circ$E and 
	$120^\circ$E. The dynamo simulation at $Ra^C =2.75
	\times 10^9$ ($\approx 180 \times Ra^C_{cr}$)
	 and $q^*=10$ shows lower longitudinal 
	variations of the Siberian lobe relative to
	the Canadian lobe (figure \ref{fig:Br_plots} (d)), 
	consistent with that in
	Earth's field.
	The Canadian lobe moves between $65^\circ$W and 
	$130^\circ$W and occasionally disappears while
	 the Siberian lobe moves between $90^\circ$E and 
	$130^\circ$E.	
	
	\begin{figure}[!htbp]
		\centering
		\subfloat[]{\includegraphics[width=0.45\textwidth, keepaspectratio]{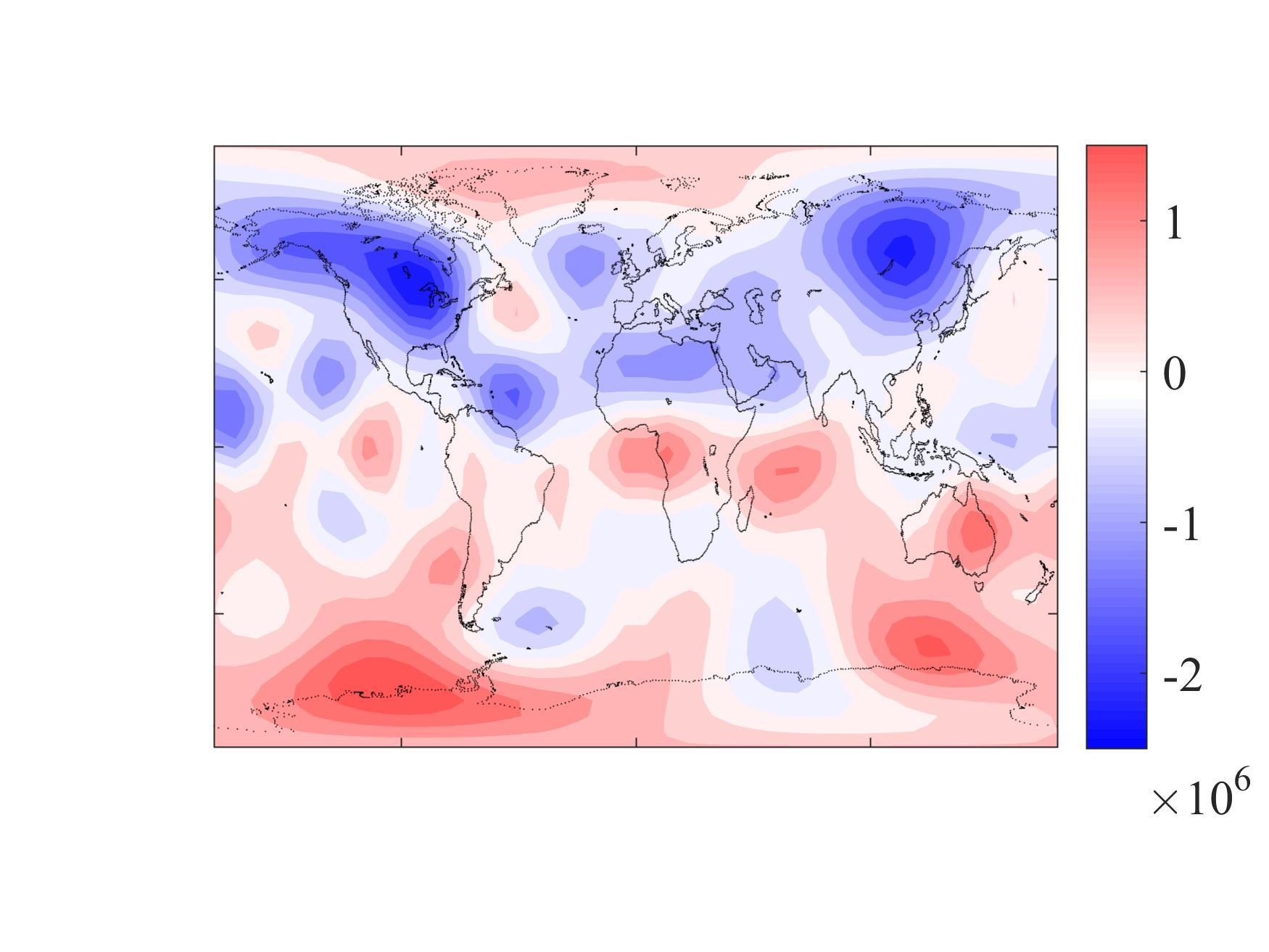}}
		\subfloat[]{
			\includegraphics[width=0.45\textwidth, keepaspectratio]{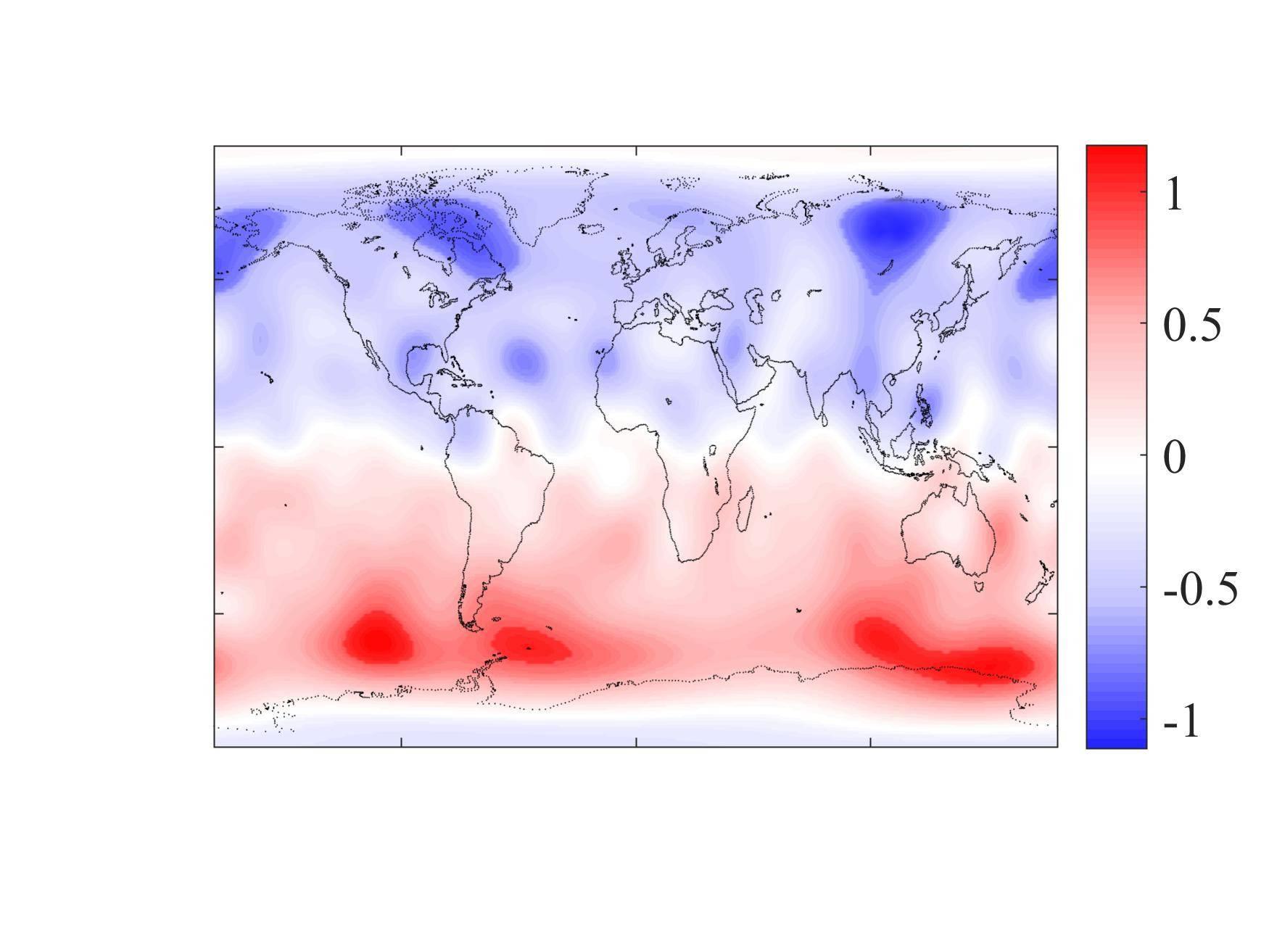}} \\
	\vspace{-1cm}
		\subfloat[]{
			\includegraphics[width=0.45\textwidth, keepaspectratio]{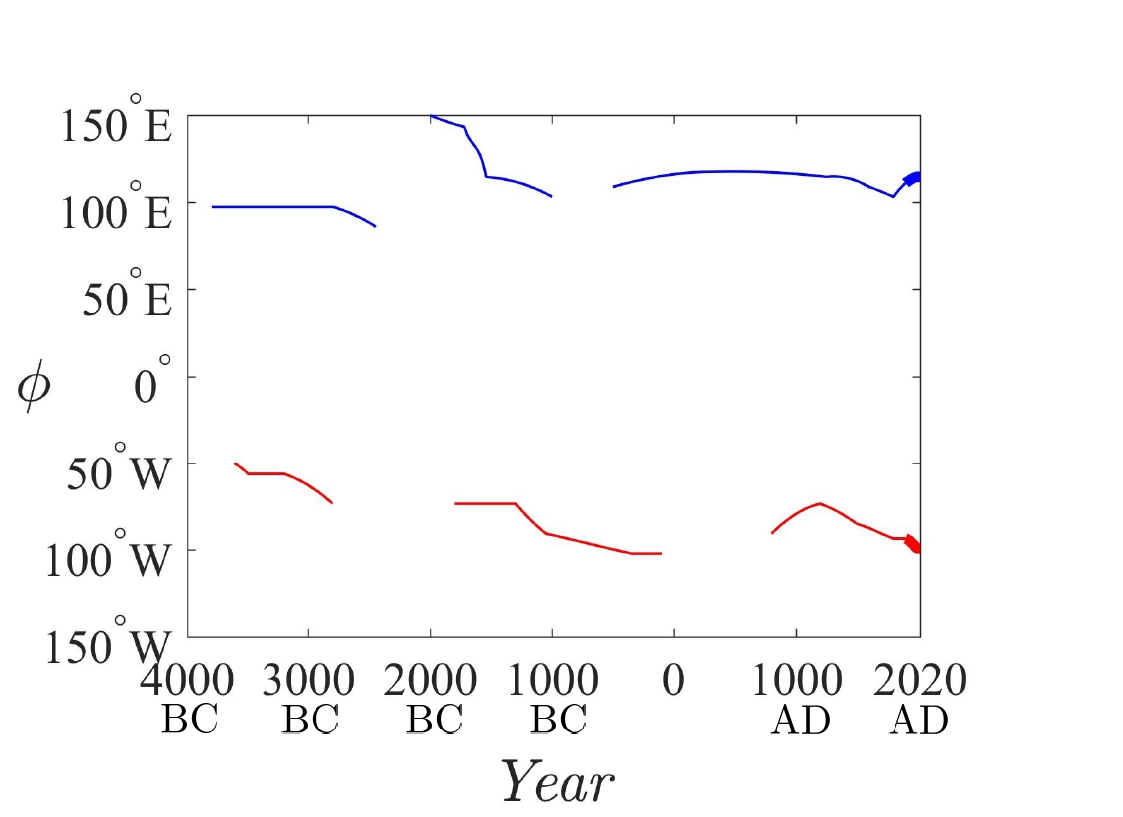}}
            \subfloat[]{
         	\includegraphics[width=0.45\textwidth, keepaspectratio]{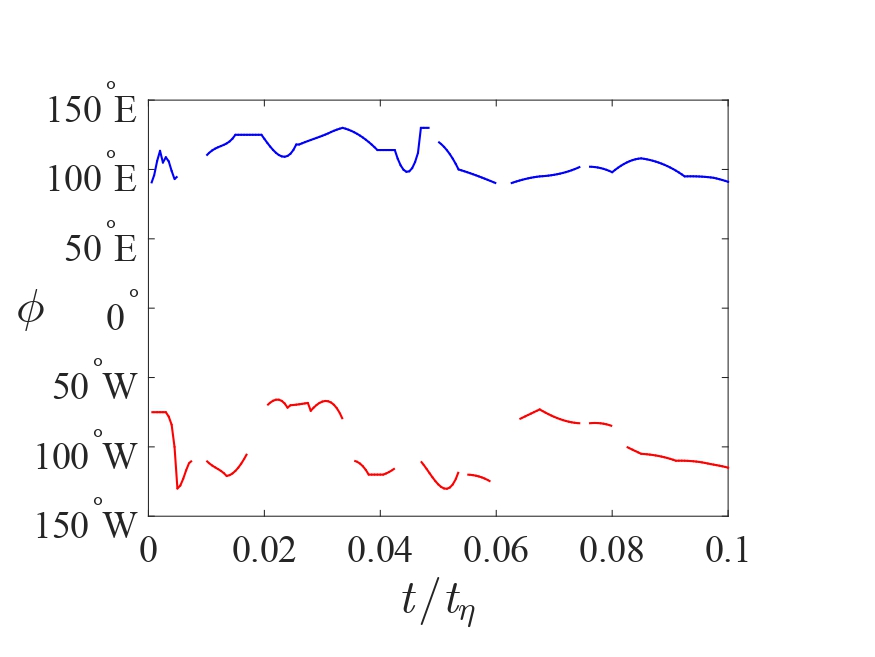}}
	\caption{Time-averaged $B_{r}$ at the CMB for (a) CALS10k.2 model for the years 1590-1990 AD and in (b) $B_r$ in 
the dynamo. (c), (d) Time series of the 
longitude of the peak value of the high-latitude flux
patches in the Northern hemisphere for the 
CALS10k.2+ CHAOS-7.12 field model 
and the dynamo respectively. The dynamo parameters are 			
		$Ra^T=4.30 \times 10^6$ 
		($\approx Ra^T_{cr}$), $Ra^C=2.75 \times 10^9$
		($\approx 180 \times Ra^C_{cr}$), $ q^*=10$, 
		with $Pr=0.1$, $Sc=1$, $Pm=1$, and $E=2 \times 10^{-6}$. 
The field at the CMB is plotted for the 
spherical harmonic degree range $ 1 \leq l \leq 10$. }
	\label{fig:Br_plots}
	\end{figure}

	\section{Discussion}
	\label{disc}

	The study shows that a thermally heterogeneous 
CMB  induces a substantial variation of heat flux at
the ICB through the medium of outer core convection.
A two-component convective dynamo model, where thermal
convection is near critical and compositional convection is
strongly supercritical, produces inner core heterogeneity 
in the strongly driven regime of present-day Earth.
The temperature profile that models secular cooling
ensures that the CMB thermal heterogeneity is transmitted
to the ICB. The precise distribution of heat flux at the ICB, on the
other hand, is determined by the strength of
compositional buoyancy, measured by
$Ra^C$. High values of $Ra^C$ ($\sim 10^2 \times Ra^C_{cr}$ in this study)
produces a long-lived downwelling beneath Asia and time-varying convection
beneath America, which in turn results in a higher 
average freezing rate and seismic
P-wave velocity in the eastern hemisphere
 relative to the western hemisphere. Since laboratory models
 of core convection \citep{sahoo2020b} indicate that the CMB heterogeneity would
 penetrate as far as the ICB when convection originates in
 close proximity to the boundary anomalies,
 the success of the two-component model rests on the
 temperature profile rather that the ratio of thermal to
 compositional diffusivities.

For moderate CMB heterogeneity, measured by $q^*$ of $O(1)$, the
ICB heat flux
pattern is approximately two-fold, with approximately
equal peaks of heat flux in the east and west and no
inward (negative) heat flux (figure \ref{fig:Q_qcomp}(b)). 
For a relatively large CMB heterogeneity of $q^*$
of $O(10)$, heat flux peak beneath Asia dominates that beneath America,
and regions of inward heat flux develop strongly beneath the Pacific and
weakly beneath Africa (figure \ref{fig:Q_qcomp}(c)). 
This two-fold pattern with dominant heat flux beneath Asia
is supported by observations
\citep[e.g.][]{cormier2013,tkalvcic2024}. 
 The departure
	from the hemispherical ($m=1$) heterogeneity points towards the 
	influence of the CMB on the freezing
	rate heterogeneity of the inner core.		 
	Increasing $q^*$ to higher values necessitates a higher
	$Ra^C$ to obtain comparable average heat flux differences
	at the ICB. 
	Increasing $Ra^C$ to high 
	values for a given $q^*$
	causes the inward heat flux to weaken and the heat flux averages 
	in the hemispheres to equalize again 
	(figure \ref{fig:Q_qcomp}(d)) as
	the convection in the eastern hemisphere also 
	becomes time-varying.  	The values
	of ($q^*$, $Ra^C/Ra^C_{cr}$) proposed in this study
	are conservative lower bounds for Earth.
	An
	upper bound of $Ra^C/Ra^C_{cr}$ would be $O(10^3)$,
	at which strong anticyclonic polar vortices must exist below
	the threshold for polarity reversals in the inertia-less
	core \citep{deb2024}. This may in turn place an upper bound
	for $q^*$.

Many studies \citep{loper1977,wong2018} 
describe the F-layer at the
	 base of the outer core as a two-phase, 
	 two-component slurry, where solid particles freeze throughout
	 the layer and sink under gravity while the light elements migrate
	 to the outer core without disturbing the layer. 
	 This study, on the other hand, considers
	 the liquid produced by localized melting
	 of the inner core to be the source of the 
	 F-layer. 
Although preferential melting in one hemisphere
can result from convective translation of the 
inner core \citep{alboussiere2010,deguen2014f},
here we propose that the lower mantle, through
outer core convection, causes regional melting of 
the inner core, in line with \cite{gubbins2011}. 
The liquid produced by melting then spreads over 
the surface of the inner core. Since the magnetic damping of
this lateral flow happens on a shorter time scale than
viscous damping, the build-up of melt over regions of negative
heat flux would be higher in the presence of the magnetic
field. Thus, in the steady state,
larger layer heights $\overline{h} \sim 200$ km
are realizable with
magnetic damping than with viscous damping.

	 The parameter space of ($q^*$, $Ra^C/Ra^C_{cr}$)
	 that gives long-lived convection in the east
	 and time-varying convection in the west satisfies
	 multiple observational constraints. First, the variability
	 of high-latitude magnetic flux in the east is markedly 
	 lower than that in the west. Second, the average
	 seismic velocity at
	 the top of the inner core is higher in the
	 east than that in the west,
	 with the peak velocity beneath equatorial East Asia.
	 Finally, the stratified layer at the base of the outer core,
	 which is fed by the mass flux from
	 regional melting of the inner core 
	 and magnetically damped, attains a steady-state
	 height of $\sim$ 200 km. All of the above factors
	 indicate that the lower mantle has a dominant effect
	 on inner core heterogeneity.
	 	
\section*{Acknowledgements}	

This study was supported by Research Grant 
CRG/2021/002486 awarded
by the Science and Engineering Research Board (India).
The computations were performed on 
\emph{Param Pravega}, the supercomputer at the
Indian Institute of Science, Bangalore.

\appendix
	\section{Dynamo driven by one-component convection} \label{sec:codensity}
	
	The non-dimensional MHD equations for the velocity
	$\bm{u}$, magnetic field $\bm{B}$ and temperature $T$
	are given by
	
	\begin{equation}
		\begin{split}
			\dfrac{E}{Pm}\left( \dfrac{\partial  \bm u}{\partial t} +
			(\bm \nabla \bm \times  \bm u ) \times \bm u\right)
			+ \hat{\bm z} \times \bm u = & -\mathbf \nabla p^{*}
			+ (\bm \nabla \bm \times  \bm B ) \times \bm B 
			\\+ &Pm Pr^{-1} E Ra T \bm r + E \nabla^{2} \bm u,
		\end{split}
	\end{equation}
	
	\begin{equation}
		\dfrac{\partial T}{\partial t} + (\bm u.\bm \nabla)T= Pm Pr^{-1}\nabla^{2}T,
	\end{equation}

	\begin{equation}
		\dfrac{\partial \bm B}{\partial t} =
		\bm \nabla \bm \times (\bm u \bm \times \bm B)
		+  \nabla^{2} \bm B,
	\end{equation}
	
	\begin{equation}
		\bm \nabla.\bm{u} = \bm \nabla.\bm{B} = 0,
	\end{equation}
where 	$p^* = p + \dfrac{1}{2} E Pm^{-1} | \bm{u}^2|$ is the
modified pressure and 
	$\hat{\bm{z}}$ = [$\cos \theta$,- $\sin \theta$ ,0]. The
	dimensionless parameters in the above equations are as
	defined in Section 2.1 of the paper.	
	
The basic state temmperature profile is obtained by solving 

\begin{equation}
	Pm Pr^{-1} \,\dfrac{1}{r^2}\dfrac{\partial}{\partial r}\left( r^2 \dfrac{\partial T_0}{\partial r}\right) - S_{i}=0,  	
\end{equation} 
which gives
\begin{equation}  \label{eq:thermprofile}
	T_0= Pr Pm^{-1} \frac{S_i r^2}{6} +\frac{C_1}{r} +C_2. 
\end{equation}	
The boundary conditions are chosen as
\begin{align}     
	& T_0 =0  \quad &\text{at} \quad  r=r_{i}, \\        
	& \dfrac{\partial T_0}{\partial r}= -0.01  \quad &\text{at}  \quad  r=r_{o}.
\end{align}	
Using $S_{i}=1$ and $Pm Pr^{-1}=1$, we obtain
\begin{equation}  \label{eq:thermprofile1}
	T_0= \frac{r^2}{6} +\frac{1.236}{r} - 4.32,
\end{equation}
which is the combination of basal heating and a heat sink
\citep{sreenivasan2008}.

\bibliography{mac4}

\end{document}